\documentclass[12pt]{article}

\usepackage{graphicx}
\usepackage{amsmath}
\usepackage{amsthm}
\usepackage{amsfonts}
\usepackage{mathrsfs}
\usepackage[utf8]{inputenc}
\usepackage{latexsym}

\newtheorem{teor}{Theorem}[section]
\newtheorem{lema}{Lemma}[section]
\newtheorem{cor}{Corollary}[section]
\newtheorem{obs}{Remark}[section]
\newtheorem{defin}{Definition}[section]
\newtheorem{exem}{Example}[section]

\newtheorem{prop}{Proposition}[section]

\newfont{\Mb}{msbm10}
\newcommand{\C}{\mbox{\Mb\symbol{67}}}
\newcommand{\R}{\mbox{\Mb\symbol{82}}}
\newcommand{\N}{\mbox{\Mb\symbol{78}}}

\begin{document}

\hspace\parindent
\thispagestyle{empty}

\bigskip
\bigskip
\bigskip
\begin{center}
{\LARGE \bf A linear probabilistic method}
\end{center}
\begin{center}
{\LARGE \bf to compute integrating factors}
\end{center}
\begin{center}
{\LARGE \bf using nonlocal symmetries}
\end{center}

\bigskip

\begin{center}
{\large
$^a$I. Deme, $^a$L.G.S. Duarte and $^a$L.A.C.P. da Mota \footnote{E-mails: idrissdeme@yahoo.fr, lgsduarte@gmail.com and lacpdamota@gmail.com}
}

\end{center}

\bigskip
\centerline{\it $^a$ Universidade do Estado do Rio de Janeiro,}
\centerline{\it Instituto de F\'{\i}sica, Depto. de F\'{\i}sica Te\'orica,}
\centerline{\it 20559-900 Rio de Janeiro -- RJ, Brazil}

\bigskip\bigskip
\bigskip
\bigskip

\abstract{Here we present an efficient method for finding and using a nonlocal symmetry admitted by a rational second order ordinary differential equation (rational 2ODE) in order to determine a Darboux integrating factor. The central idea consists in using a nonlocal symmetry of the 2ODE to define three polynomial vector fields (in $\R^2$), which `share' the first integral of the rational 2ODE. These plane polynomial vector fields can be used to construct a linear procedure using a fast probabilistic algorithm to determine an integrating factor for the rational 2ODE. The main advantages of the proposed method are: the obtaining of the nonlocal symmetry is algorithmic and very efficient and, furthermore, its use to find an integrating factor is a sequence of fast linear processes.}

\bigskip
\bigskip
\bigskip
\bigskip
\bigskip
\bigskip



\newpage

%%%%%%%%%%%%%%%%%%%%%%%%%%%%%%%%%%%%%%%%%%%%%%%%%%%%%%%%%%%%%%%%%%%%%
%%%%%%%%%%%%%%%%%%%%%%%%%%%%%%%%%%%%%%%%%%%%%%%%%%%%%%%%%%%%%%%%%%%%%
\section{Introduction}
\label{intro}

When the topic is the search for Liouvillian first integrals (LFIs) of rational second order ordinary differential equations (rational 2ODEs), the Lie and Darboux methods are as a rule the most comprehensive and, in general, the most efficient as well. However, rational 2ODEs not admiting Lie point symmetries and presenting integrating factors formed by relatively high degree Darboux polynomials (DPs) can make this task very complicated or even unfeasible in practice. In an attempt to cover these cases, J. Avellar {\it et al} developed, in \cite{Noscpc2019}, a method to determine LFIs of rational 2ODEs (called the S-function method) based on another framework: the central idea is to construct a rational first order ordinary differential equation (rational 1ODE) such that the function defining its general solution in implicit form is precisely the LFI of the original 2ODE\footnote{In other words, instead of trying to find a LFI of a rational 2ODE, the idea is to construct (and solving) a rational 1ODE associated with the original 2ODE.}. Mathematically, the determination of the S-function is equivalent to determining a nonlocal symmetry (or a $\lambda$-symmetry)\footnote{For a summary of approaches that use nonlocal or $\lambda$-symmetries, see \cite{AbrGuo,AbrGuo2,AbrGovLea,Abr,GovLea,AdaMah,GanBru,GanBruSen} (for nonlocal symmetries) and \cite {MurRom,MurRom2,MurRom3,MurRom4} (for $\lambda$-symmetries).}. The S-function method differs from others in that it uses an algebraic procedure for finding the S-function (see, for example, \cite{Nosjpa2001,Nosamc2007,Nosjmp2009,Nosjpa2010}), 
instead of the usually applied process that consists in splitting the partial differential equation (PDE) that establishes the symmetry condition (commonly called {\em determining equation}) into an overdetermined system of PDEs. Some very interesting approaches managed to circumvent this problem for a wide variety of cases, such as the $\lambda$-symmetry method. Nevertheless, for the general case, i.e., when the $\lambda$-symmetry depends on $(x,y,y')$, the problem remains. For these cases, the S-function method (which was already a good option -- see \cite{Noscpc2019}) was made faster in \cite{Noscpc2024} and it has become an excellent alternative for dealing with 2ODEs where the Lie and Darboux methods had problems, i.e., 2ODEs with complicated symmetries and with high degree DPs in the integrating factor.

Despite all this, the use of the nonlocal symmetry (or, equivalently, of the $\lambda$-symmetry or the S-function) in order to find a LFI is the same: the final part of the method consists of solving a 1ODE which, in practice, can be quite complicated. But, as we will show, there is another way of using the nonlocal symmetry in which it is possible to build a linear probabilistic procedure -- that can be computationally implemented in a very practical way -- to determine an integrating factor for the 2ODE.

In this article we construct a linear probabilistic algorithm to compute the Darboux polynomials (DPs) present in the integrating factor of a rational 2ODE\footnote{The problem of determining, from a polynomial vector field, a Darboux integrating factor (or determining that such a factor does not exist) is old and difficult. Since a Darboux integrating factor is formed by Darboux polynomials (DPs), a previous problem consists of, given a polynomial vector field, determining the invariant algebraic curves/hypersurfaces (that define the DPs) of the vector field, or ensuring that such curves/hypersurfaces do not exist. For an overview see \cite{PreSin,Sin,Sch,Chr,Nosjpa2002-1,Nosjpa2002-2,Noscpc2002,Nosjcam2005,LliZha,ChLlPaWa4,Zha,FeGi,Che,BoChClWe,FeGa,Dem,FeGaMo,
ChCo,ChLlPaWa5,Nosjde2021} and references therein.}. 
The main results can be found in section \ref{lpffi} and the paper has the following structure: 

\begin{enumerate}
\item The first section corresponds to this introduction.

\item In the second section, we present (briefly) an improved version of the S-function method presented in \cite{Noscpc2019,Noscpc2024} to determine a non-local symmetry admited by a rational 2ODE that presents a LFI. We also rewrote certain results in a more formal way and constructed a notation more suitable for the presentation that follows. We finish the section by presenting some examples.

\item In the third, we develop a method that uses the nonlocal symmetry to linearly determine a Darboux integrating factor (without the need to solve any differential equation) using a probabilistic algorithm. 
\begin{itemize}
\item In the first subsection, we show some important results: 
\begin{enumerate} 
\item A rational 2ODE with a LFI (belonging to a large class of Liouvillian functions) has a Darboux integrating factor $R$. 
\item This 2ODE admits a symmetry $\mathfrak{S}$ (That can be computed with the algebraic algorithm presented in the second section). 
\item From the symmetry $\mathfrak{S}$, we can build three associated plane polynomial vector fields $\mathfrak{X}_i$, ($i \in \{1,2,3\}$) such that the LFI $I$ of the 2ODE is also a first integral for vector fields $\mathfrak{X}_i$ (i.e., $\mathfrak{X}_i(I)=0$).
\end{enumerate}

\item In the second, in the spirit of the idea presented in \cite{Nosjde2021}, we define another three polynomial vector fields ${\cal X}_i$, ($i \in \{1,2,3\}$) associated with $\mathfrak{X}_i$, such that their first integral ${\cal I}$ is the Darboux integrating factor $R$ of the rational 2ODE, i.e., ${\cal X}_i(R)=0$. We show how to use the vector fields ${\cal X}_i$ to find a Darboux integrating factor $R$ with a linear setting. 

\item In the third, we propose a probabilistic procedure to determine a LFI for the rational 2ODE. To make the algorithm clear, we present a worked example step by step.
\end{itemize}

\item In the fourth section, we apply the procedures presented in sections \ref{fnls} and \ref{lpffi} to some rational 2ODEs and discuss its performances:

\begin{itemize}
\item In the first subsection, we apply the method to some rational 2ODEs that are `difficult' for Lie and Darboux methods (as well as for the $S$-function method) and we analyze the performance of the algorithms.
\item In the second, we comment on the advantages and disadvantages of the method and make some considerations about possible developments and extensions of it.
\end{itemize}

\end{enumerate}

%%%%%%%%%%%%%%%%%%%%%%%%%%%%%%%%%%%%%%%%%%%%%%%%%%%%%%%%%%%%%%%%%%%%%
%%%%%%%%%%%%%%%%%%%%%%%%%%%%%%%%%%%%%%%%%%%%%%%%%%%%%%%%%%%%%%%%%%%%%
\section{Finding a nonlocal symmetry}
\label{fnls}

In this section, we present an algorithm to determine a symmetry of a rational 2ODE that has a LFI belonging to the class $L_S$ defined below. The procedure consists of an improvement of the method developed in \cite{Noscpc2019} and, in addition to being more comprehensive, it is also more efficient than the original S-function method. First, we will make some definitions and establish some results in a context more appropriate to what we will present.

\medskip
\noindent
Consider a rational 2ODE 
\begin{equation}
\label{r2ode}
z' = {\frac{M(x,y,z)}{N(x,y,z)}} = \phi(x,y,z), \,\,\,\, (z \equiv y'),
\end{equation}
where $M$ and $N$ are coprime polynomials in $\C[x,y,z]$. A first integral $I$ of the 2ODE (\ref{r2ode}) is a function that is constant over the solutions of (\ref{r2ode}).

\begin{defin}
\label{mathX}
Let $L$ be a Liouvillian field extention\footnote{For a formal definition of Liouvillian field extention see \cite{Dav}.} of $\C(x,y,z)$. A function $I(x,y,z) \in L$ is said to be a {\bf Liouvillian first integral} ({\bf LFI}) of the rational {\em 2ODE (\ref{r2ode})} if $\mathfrak{X}(I)=0$, where $\mathfrak{X} \equiv N\, \partial_x + z\,N\, \partial_y + M\, \partial_z$ is the {\bf polynomial vector field associated} (or {\bf Darboux operator associated}) with the {\em 2ODE (\ref{r2ode})}.
\end{defin}

\begin{defin}
Let $p(x,y,z)\,\in\, \C[x,y,z]$. The polynomial $p$ is said to be a {\bf Darboux polynomial} ({\bf DP}) of the vector field $\mathfrak{X}$ if $\mathfrak{X}(p)=q\,p$, where $q$ is a polynomial in $\C[x,y,z]$ called {\bf cofactor} of $p$.
\end{defin}

\begin{defin}
\label{ourlfi}
Let $I(x,y,z)$ be a LFI of the rational {\em 2ODE (\ref{r2ode})} and consider that its derivatives can be written as $\partial_i(I) = R P_i$ ($i \in \{1,2,3\},\,x_1\equiv x,\,x_2\equiv y,\,x_3 \equiv z$) where $P_i$ are coprime polynomials in $\C[x,y,z]$ and $R$ is a Liouvillian function of $(x,y,z)$. In this case we say that $I$ is a member of the set \mbox{\boldmath $L_S$} and that $R$ is an {\bf integrating factor} associated with $I$.
\end{defin}

\begin{defin}
\label{omg}
Let $\omega$ be a polynomial 1-form defined by $\omega \equiv \sum_i P_i\, dx_i$. A function $G$ is said to be an {\bf integrating factor} for the 1-form $\omega$ if the 1-form $G\,\omega$ is exact.
\end{defin}

\begin{obs}
\label{ourif}
From definitions {\em \ref{ourlfi}} and {\em \ref{omg}}, we have that $R$ is an integrating factor for the 1-form $\omega$.
\end{obs}

\begin{defin}
\label{ourvfi}
Let $I \in L_S$ be as above. We define $\mathfrak{I} \equiv  \sum_i P_i\,\partial_i$.
\end{defin}

\begin{obs}
\label{idop}
The condition $\mathfrak{X}(I)=0$ is equivalent to $\langle \mathfrak{X}|\mathfrak{I} \rangle = N\,P_1+z\,N\,P_2+M\,P_3 = 0$.$\,$\footnote{In what follows, the operators $\,\nabla(.),\, \langle \nabla | . \rangle,\, \nabla \wedge .\,$ stand for {\bf grad, div, culr}, respectively.}
\end{obs}

\medskip

\noindent
The vector field $\mathfrak{I}$ can be used to build a symmetry for the 2ODE (\ref{r2ode}):

\begin{teor} 
\label{teosymm}
Let $I \in L_S$ be a first integral of the rational {\em 2ODE (\ref{r2ode})} as above. Then, {\em 2ODE (\ref{r2ode})} admits a symmetry given by
\begin{equation}
\label{symmev}
\mathfrak{S} = {\rm \large{e}}^{\int_x(-P_2/P_3)}\,\partial_y,
\end{equation}
where $\, \int_x \,\,\mbox{\rm is the inverse operator of}\,\, D_x\,,\,\left(D_x \equiv \frac{\mathfrak{X}}{N}\right)$.
\end{teor}

\medskip

\noindent
{\it Proof.}  If $\mathfrak{S} = {\rm \large{e}}^{\int_x(-P_2/P_3)}\,\partial_y$ is a symmetry (in evolutionary form) of the rational 2ODE (\ref{r2ode}), then its first prolongation is given by
\begin{equation}
\mathfrak{S}^{(1)} = {\rm \large{e}}^{\int_x(-P_2/P_3)}\,\partial_y+D_x\left({\rm \large{e}}^{\int_x(-P_2/P_3)}\right)\,\partial_z = {\rm \large{e}}^{\int_x(-P_2/P_3)}\!\left(\partial_y -(P_2/P_3)\,\partial_z \right)\!, \nonumber
\end{equation}
and, so, $\,\mathfrak{S}^{(1)}(I) = {\rm \large{e}}^{\int_x(-P_2/P_3)} \left(I_y - (P_2/P_3)\,I_z \right)$. Since $I \in L_S$, $I_y = R\,P_2$ and $I_z = R\,P_3$. Then $I_y -\frac{P_2}{P_3}\,I_z = R\,P_2 -\frac{P_2}{P_3}\,R\,P_3 = 0 \,\,\,\Rightarrow \,\,\, \,\mathfrak{S}^{(1)}(I) =0$. $\,\,\,\Box$

\bigskip

Therefore, if we find the vector field $\,\mathfrak{I}$ (or at least two of its components) then we have the symmetry $\mathfrak{S}$. In order to find it we will first show that the conditions the polynomials $P_i$ must satisfy can be written as first order partial differential equations (1PDEs) of Riccati type (that will be an important step in our method).

\begin{lema}
\label{peli}
Let $I \in L_S$ and $\mathfrak{I}$ be as above. Then the vector field defined by $\mathfrak{L} \equiv \mathbf{\nabla} \wedge \mathfrak{I}$ obeys the condition $\langle \mathfrak{L},\mathfrak{I} \rangle = 0$.
\end{lema}

\noindent
{\it Proof.}  From the hypotheses of the lemma $\,I_i = R\,P_i$. So, we can write $\mathfrak{I}$ as $\frac{\mathbf{\nabla}(I)}{R}$ implying that 

\noindent
$ \mathfrak{L} = \mathbf{\nabla} \wedge \mathfrak{I} = \displaystyle{\frac{1}{R}\,\underbrace{\nabla \wedge \nabla(I)}_{=\, 0} + \nabla\left(\frac{1}{R}\right) \wedge \nabla(I)}$.

\noindent
So, $\langle \mathfrak{L},\mathfrak{I} \rangle = \langle \nabla\left(\frac{1}{R}\right) \wedge \nabla(I),\frac{\mathbf{\nabla}(I)}{R} \rangle = 0$. $\,\,\,\Box$

\medskip

\begin{lema} 
\label{XxLigualTI}
Let $I \in L_S$, $\mathfrak{X}$, $\mathfrak{I}$ and $\mathfrak{L}$ be defined as above. Then $\,\,\,\mathfrak{X} \wedge \mathfrak{L} = \frac{\mathfrak{X}(R)}{R}\,\mathfrak{I}$.
\end{lema}

\noindent
{\it Proof.}  From lemma \ref{peli} we have that $\mathfrak{L} = \nabla\left(\frac{1}{R}\right) \wedge \nabla(I)$. So\footnote{Using the identity $A \wedge (B \wedge C) = \langle A,C \rangle\,B - \langle A,B \rangle\,C$.}

\medskip

\noindent
$\mathfrak{X} \wedge \mathfrak{L} = \mathfrak{X} \wedge \left(\nabla\left(\frac{1}{R}\right) \wedge \nabla(I) \right) = 
\underbrace{\langle \mathfrak{X}|\nabla(I) \rangle}_{\mathfrak{X}(I)}\, \nabla\left(\frac{1}{R}\right) - \langle \mathfrak{X}|\nabla\left(\frac{1}{R}\right) \rangle\,\nabla(I)$. 

\medskip

\noindent
Since $\mathfrak{X}(I)=0$, $\,\,\mathfrak{X} \wedge \mathfrak{L} = - \langle \mathfrak{X}|-\frac{\nabla\left(R\right)}{R^2} \rangle\,\nabla(I) = \frac{\mathfrak{X}(R)}{R}\,\frac{\nabla(I)}{R} = \frac{\mathfrak{X}(R)}{R}\,\mathfrak{I}. \,\,\,\Box$

\medskip

\begin{obs}
What was stated in lemma {\em \ref{peli}} is the well known integrability condition $\,\omega \wedge d\omega = 0\,$ for the 1-form $\omega$. The point of writing the lemmas {\em \ref{peli}} and {\em \ref{XxLigualTI}} using the language of vector fields is simply because it is more suitable to obtain some algebraic conclusions that we are going to present.
\end{obs}

\begin{prop}
\label{eqtnp}
Let $I \in L_S$, $\mathfrak{X}$, $\mathfrak{I}$ be as above. Then the polynomials $\,P_2,\,P_3\,$ obey the following equations:
\begin{eqnarray}
\label{eq1pdeP}
&& P_2\,\frac{\mathfrak{X}\left(R\right)}{R} + N\,\partial_y\!\left(\frac{M}{N}\right)\,P_3 + \mathfrak{X}\left(P_2\right)=0, \\ [3mm] 
\label{eq1pdeN}
&& P_3\,\frac{\mathfrak{X}\left(R\right)}{R} + N\,P_2 + N\,\partial_z\!\left(\frac{M}{N}\right)\,P_3 + \mathfrak{X}\left(P_3\right)=0.
\end{eqnarray}
\end{prop}

\noindent
{\it Proof.} In order to avoid tedious calculations, we will only indicate the process for obtaining the desired result: From the hypotheses of the theorem we have that $\,\langle \mathfrak{X}|\mathfrak{I} \rangle =  N\,P_1+z\,N\,P_2+M\,P_3 = 0$. Solving it for $\,P_1,$ and substituting the result in $\,\,\mathfrak{X} \wedge \mathfrak{L} = \frac{\mathfrak{X}(R)}{R}\,\mathfrak{I}\,$ (lemma \ref{XxLigualTI}) one obtains (after rearranging terms) the conditions expressed by 1PDEs (\ref{eq1pdeP},\ref{eq1pdeN}). $\,\,\,\Box$

\medskip

\begin{obs}
\label{solforPeN}
There are analogous conditions for the other two pairs $\,\{P_1,P_3\}\,$ and $\,\{P_1,P_2\}\,$ that result from solving the equation $N\,P_1+z\,N\,P_2+M\,P_3 = 0$ for $P_2$ and $P_3$, respectively, and substituting the results into the equation $\,\,\mathfrak{X} \wedge \mathfrak{L} = \frac{\mathfrak{X}(R)}{R}\,\mathfrak{I}\,$. 
\end{obs}

\begin{obs}
\label{psneq}
We can eliminate the term $\frac{\mathfrak{X}\left(R\right)}{R}$ from equations (\ref{eq1pdeP},\ref{eq1pdeN}) and obtain an equation for the polynomials $P_2$ and $P_3$. Analogously, we can eliminate it from the pairs of equations for $\,\{P_1,P_3\}\,$ and $\,\{P_1,P_2\}\,$, obtaining another two equations (see theorem {\em \ref{eqsss}} bellow).
\end{obs}

\begin{teor}
\label{eqsss}
Let $I \in L_S$, $\mathfrak{X}$, $\mathfrak{I}$ be defined as above. Then the following equations hold:
\begin{equation}
\label{eq3s}
\mathfrak{X}\left(\!\frac{P_j}{P_k}\!\right)\!=\!F_i\,\partial_k\!\left(\!\frac{F_j}{F_i}\!\right)\left(\!\frac{P_j}{P_k}\!\right)^2\!+F_i\left(\!\partial_k\!\left(\!\frac{F_k}{F_i}\!\right)-\partial_j\!\left(\!\frac{F_j}{F_i}\!\right)\right)\left(\!\frac{P_j}{P_k}\!\right) - F_i\,\partial_j\!\left(\!\frac{F_k}{F_i}\!\right),
\end{equation}
where $\,F_1=N,\,F_2=z\,N,\,F_3=M$, and $(i,j,k)$ is any permutation of the set $\{1,2,3\}$ (i.e., there is no summation over the repeated indices in equation {\em(\ref{eq3s})}).
\end{teor}

\medskip

\noindent
{\it Proof.} From the hypotheses of the theorem we have the following conditions: $\,\langle \mathfrak{X}|\mathfrak{I} \rangle = \langle \mathfrak{L}|\mathfrak{I} \rangle = 0$. As in the proof of theorem \ref{eqtnp} (to avoid tedious calculations), we only indicate the path: solve the equation $\,\langle \mathfrak{X}|\mathfrak{I} \rangle = 0$ for one of the polynomials $\,P_i\,$ and substitute the solution in the equation $\langle \mathfrak{L}|\mathfrak{I} \rangle = 0$. We obtain (depending on which polynomial we choose) the condition expressed by (\ref{eq3s}) for the two other polynomial functions. $\,\,\,\Box$

\medskip

\begin{obs}
Equations (8) are quadratic in the rational functions $\left(\!\frac{P_j}{P_k}\!\right)$, making their resolution a difficult computational problem. 
\end{obs}

\noindent 
To avoid the difficulty pointed out in the above observation, we can make use of the following result:

\begin{prop}
\label{Nknown}
Let $I \in L_S$ and $\mathfrak{I}$ defined as above. If $P_3$ and $P_1+z\,P_2$ are coprime polynomials, then $P_3=N$.
\end{prop}

\medskip

\noindent
{\it Proof.}
From the hypotheses, $I_i = R\,P_i$ and therefore
\begin{equation}
\label{nin0}
\phi(x,y,z)=-\frac{I_x+z\,I_y}{I_z}=-\frac{R\,P_1+z\,R\,P_2}{R\,P_3}=-\frac{P_1+z\,P_2}{P_3}=\frac{M}{N}.
\end{equation}
Since $P_3$ and $P_1+z\,P_2$ are coprime, we have that $P_3=N$. $\,\,\,\Box$
\medskip

\begin{obs}
The only case where $P_3 \neq N$ can occur is if the polynomials $P_3$ and $P_1+z\,P_2$ have a non-constant polynomial factor in common. We will deal with this `degenerate' case in another paper.
\end{obs}

\begin{cor}
\label{eqXP}
If the assumptions of proposition {\em \ref{Nknown}} are fulfilled, then $P_2$  obey the equation
\begin{equation}
\label{eqxp}
\mathfrak{X}\left(P_2\right)={P_2}^2+\left(N_x+z\,N_y+M_z\right)\,P_2+M\,N_y-M_y\,N \,.
\end{equation}
\end{cor}

\medskip

\noindent
{\it Proof.}
Choosing $i=1,\,j=2,\,k=3$ in (\ref{eq3s}), we have that
\begin{equation}
\label{eqSpsn}
\mathfrak{X}\!\left(\!\frac{P_2}{N}\!\right)=N\,\partial_z\!\left(\frac{z\,N}{N}\right)\!\left(\frac{P_2}{N}\right)^2+N\left(\partial_z\!\left(\!\frac{M}{N}\!\right)-\partial_y\!\left(\frac{z\,N}{N}\right)\!\right)\! \left(\frac{P_2}{N}\right) - N\,\partial_y\!\left(\!\frac{M}{N}\!\right)\!.
\end{equation}
Multiplying both sides of (\ref{eqSpsn}) by ${N}^2$ and isolating the term $\mathfrak{X}\left(P_2\right)$ on the left hand side we arrive at equation (\ref{eqxp}). $\,\,\,\Box$

\bigskip

\noindent
The equation (\ref{eqxp}) already indicates a good algorithm for finding the nonlocal symmetry: we could construct a polynomial candidate with undetermined coefficients $p_i$, substitute it into (\ref{eqxp}), and try to solve the resulting algebraic system for the undetermined coefficients. However, the presence of the term $P^2$ implies that the algebraic system is quadratic in the $p_i$, which can be very difficult to solve (computationally speaking) if the degree of the 2ODE is relatively high. In order to make the process more efficient we will use another corollary to proposition \ref{Nknown}.

\begin{cor}
\label{theimp4}
Assume that the hypotheses of proposition {\em\ref{Nknown}} hold. If none of the monomial terms of $z\,P_2$ and $P_1$ cancel, then all the monomials of $P_2$ are in $M_z$.
\end{cor}

\noindent
{\it proof.}
The conclusion follows directly from the fact that $\, M = - (P_1+z\,P_2)$. $\,\,\,\Box$

\bigskip

\noindent
Therefore, if the conditions of corollary \ref{theimp4} hold, we can just use the monomials of $M_z$ to construct the candidate $P_c$. 

\medskip

\noindent
{\bf Algorithm {\it NLS} (sketch)}:
\begin{enumerate}
\item Select the monomials from $M_z$: $[mon_1,mon_2,...,mon_k]$.
\item Construct a polynomial candidate $P_c$ with undetermined coefficients $p_i$: $P_c := {\sum^k}_{\!\!\! i=1} \,p_i\,mon_i$.
\item Substitute $P_c$ in equation $\mathfrak{X}\left(P_2\right) - {P_2}^2 - (N_x + z\,N_y +M_z)\,P_2-M\,N_y+M_yN=0$ and collect the resulting polynomial equation in the variables $(x, y,z)$ obtaining a set of equations for the $p_i$.
\item Solve this set of equations and substitute the solution on the candidate $P_c$ to
obtain $P_2$.
\item The symmetry is $\mathfrak{S} = {\rm \large{e}}^{\int_x(-P_2/N)}\,\partial_y$.
\end{enumerate}

\medskip
\begin{exem}
\label{exsymm3}
\end{exem}

\noindent
Consider the 2ODE
\begin{equation}
\label{exeufiz}
z'=-{\frac {x \left( -2\,{x}^{2}y{z}^{3}+3\,x{y}^{2}{z}^{2}+{x}^{3}z-2\,x
yz+2\,{y}^{2} \right) }{{x}^{5}+{y}^{4}}}.
\end{equation}

\medskip

\noindent
Applying procedure {\it NLS} to 2ODE (\ref{exeufiz}):
\begin{itemize}
\item The monomials of $M_z$ are: $[{x}^{3}y{z}^{2},{x}^{2}{y}^{2}z,{x}^{4},{x}^{2}y]$.

\item $P_c = p_{1}\,{x}^{3}y{z}^{2}+p_{2}\,{x}^{2}{y}^{2}z+p_{3}\,{x}^{4}+p_{4}\,{x}^{2}y$.

\item From steps 3 and 4 we obtain the solution:

\noindent
$\{ p_{1}=-2,p_{2}=0,p_{3}=0,p_{4}=-2 \}$.

\noindent
Substituting it on $P_c$, we obtain: $P_2 = -2\,y{x}^{2} \left( x{z}^{2}+1 \right) $.

\end{itemize}

\medskip

\begin{obs}
\label{obsfun11}
The algorithm {\it NLS} spent 0.05 sec of CPU time and approximately 0 MB of memory consumption.\footnote{These data (referring to CPU time and memory consumption) were established in a test of the algorithm {\it NLS} implemented in Maple 17 running on a notebook with an Intel I5 processor - 4GB - 8th Gen.}
\end{obs}

\begin{obs}
\label{obsfun12}
In the case (rare) where some of the monomial terms of $z\,P_2$ and $P_1$ cancel, we can follow a strategy similar to the one used in the blow up technique (which is used in desingularization of degenerate singular points of planar vector fields --- see {\em \cite{AlFeJa}} and references therein). The idea is to perform a transformation of variables that changes the monomials of $z\,P_2$ and $P_1$ in a disproportionate way avoiding cancellation.
\end{obs}

\medskip

\noindent
Let's see how we can deal with the situation described in remark \ref{obsfun12} (in practice) in the next example:

\medskip
\begin{exem}
\label{exsymm4}
\end{exem}

\noindent
Consider the 2ODE
\begin{equation}
\label{ex2eufiz}
z' = {\frac {z \left( {x}^{2}{z}^{5}-{x}^{2}{z}^{4}+2\,xy{z}^{3}-x{z}^{4}-2
\,xy{z}^{2}-y{z}^{3}+x{z}^{2}+{y}^{2}z-2\,y{z}^{2}-{y}^{2}-yz \right) 
}{ \left( x{z}^{2}-xz+yz+y \right)  \left( x{z}^{2}-y \right) }}.
\end{equation}

\noindent
If we apply the algorithm {\it NLS} we will see that it fails to determine the symmetry. However, applying the transformation $\{x=x^2,y=y\}$ to the 2EDO, we obtain the following transformed 2ODE:
\begin{eqnarray}
z'&=&-z ( 2\,{x}^{3}{z}^{4}-{z}^{5}{x}^{2}+16\,{x}^{3}y{z}^{2}
-8\,{x}^{2}y{z}^{3}+32\,{x}^{3}{y}^{2}-8\,{x}^{3}{z}^{2}-                        \nonumber \\ [2mm] &&
16\,{x}^{2}{y}^{2}z+2\,{x}^{2}{z}^{3}+x{z}^{4}+8\,{x}^{2}yz+
16\,xy{z}^{2}+2\,y{z}^{3}+16\,x{y}^{2}+                                                     \nonumber \\ [2mm] &&
8\,z{y}^{2})/(x \left( -{z}^{2}+4\,y \right)\left( 2\,z{x}^{2}-x{z}^{2}-4\,xy-2\,yz \right))
\label{ex3-2ode3tr}
\end{eqnarray}
This time, applying the algorithm {\it NLS} to the transformed 2ODE, we get
\begin{equation}
\label{ex3eufizp}
P_2 = -{\left( x{z}^{4}+8\,xy{z}^{2}+8\,z{x}^{2}+16\,x{y}^{2}-4\,x{z}^{2}-16\,xy-8\,yz \right)\! xz}, 
\end{equation}

\begin{obs}
\label{examtimes1}
The CPU time and memory consumption to execute the procedure {\it NLS} applied to the transformed {\em 2ODE (\ref{ex3-2ode3tr})} was 0.08 sec and 0 MB approximately. 
\end{obs}

\medskip

%%%%%%%%%%%%%%%%%%%%%%%%%%%%%%%%%%%%%%%%%%%%%%%%%%%%%%%%%%%%%%%%%%%%%
%%%%%%%%%%%%%%%%%%%%%%%%%%%%%%%%%%%%%%%%%%%%%%%%%%%%%%%%%%%%%%%%%%%%%
\section{Using a nonlocal symmetry to determine a Darboux integrating factor}
\label{lpffi}

In this section we show how to use (in a new way) a nonlocal symmetry to construct a Darboux integrating factor for the 2ODE (\ref{r2ode}):
\begin{itemize}
\item In the first subsection we use the nonlocal symmetry to construct three 2D polynomial vector fields (associated with the 2ODE) such that they `share' the first integral and an integrating factor with the 2ODE.
\item In the second subsection, we show that there are another three 2D polynomial vector fields whose first integral is the Darboux integrating factor of the 2ODE. These vector fields can be found together with the Darboux polynomials present in the integrating factor of the 2ODE by solving {\bf linear systems of indeterminates}.
\item Finally, we propose a procedure (based on a probabilistic algorithm) to determine a Darboux integrating factor for the rational 2ODE and present an example.
\end{itemize}

%%%%%%%%%%%%%%%%%%%%%%%%%%%%%%%%%%%%%
\subsection{Vector fields (in $\R^2$) associated with the 2ODE}
\label{vfans}

\begin{teor} 
\label{teosymm}
Let $I \in L_S$, $\mathfrak{X}$ and $\mathfrak{I}$ be defined as above. Then, the following statements hold: 

\vspace{1mm}
\noindent
$(a)\,\,$ The polynomial vector fields defined by $\,\mathfrak{X}_i \equiv - \sum_{j,k} \epsilon_{ijk} P_j \,\partial_k$, $ (i,j,k \in \{1,2,3\})$,
admit $I$ as a first integral, i.e., $\,\,\mathfrak{X}_i(I)=0.$

\vspace{2mm}
\noindent
$(b) \,\,$ $\mathfrak{X}_i(R) = -R\, \langle \nabla | \mathfrak{X}_i \rangle.$
\end{teor}

\bigskip

\noindent
{\it Proof.} 

\noindent
$(a) \,\,$ The statement ({\it a}) follows directly from the definition:
\vspace{1mm} 

$\,\,\mathfrak{X}_i(I)= \langle \mathfrak{X}_i | \mathfrak{I} \rangle = \langle - \sum_{j,k} \epsilon_{ijk} P_j \,\partial_k | \sum_a P_a\,\partial_a \rangle = - \sum_{j,k} \epsilon_{ijk} P_j \, P_k = 0$.

\bigskip

\noindent
$(b) \,\,$ We have that $\nabla \wedge \nabla(I) = \nabla \wedge (R\,\mathfrak{I}) = \nabla(R) \wedge \mathfrak{I} + R\,\,\nabla \wedge \mathfrak{I} = 0$. 
\vspace{1mm} 

$\nabla(R) \wedge \mathfrak{I} = \sum_{j,k} \epsilon_{ijk} \partial_j(R) \, P_k \, \partial_i = - \sum_{j,k} \epsilon_{ijk} P_j\partial_k(R) \, \partial_i = \sum_i\mathfrak{X}_i(R)\, \partial_i $.
\vspace{2mm} 

$R\,\,\nabla \wedge \mathfrak{I} = R\, \sum_{j,k} \epsilon_{ijk} \partial_j(P_k) \, \partial_i = R\,\sum_i \langle \nabla | \mathfrak{X}_i \rangle \, \partial_i $. $\,\,\,\,\Box$

\begin{obs}
\label{xi0}
Note that the associated vector fields $\mathfrak{X}_i$ (see statement $(a)$ of theorem {\em \ref{teosymm}}) present $I$ as a Liouvillian first integral (the same LFI admitted by the vector field $\mathfrak{X}$). Therefore, by the results of Singer {\em \cite{Sin}} and Christopher {\em \cite{Chr}}, they admit a Darboux integrating factor. 
\end{obs}

\begin{obs}
\label{xrsrpol1}
Note also that, in view of statement $(b)$ (theorem {\em \ref{teosymm}}), the integrating factor $R$ of the vector field $\mathfrak{X}$ is also an integrating factor for the vector fields $\mathfrak{X}_i$.
\end{obs}

\medskip

\noindent
Developing what was pointed in the remarks \ref{xi0} and \ref{xrsrpol1}, we can infer the following result:

\medskip

\begin{teor}
\label{rdarb}
Let $I \in L_S$ be a first integral of the rational {\em 2ODE (\ref{r2ode})}. Then the 3D polynomial vector field $\,\mathfrak{X}$ (associated with it) has a Darboux integrating factor.
\end{teor}

\noindent
The key to proving this theorem comes from Singer and Christopher (SC) result for polynomial vector fields in the plane (see \cite{Sin,Chr,Nosjpa2002-2}): {\em The existence of a Liouvillian first integral admitted by a polynomial vector field (in the plane) implies the existence of a Darboux integrating factor.}

\medskip

\noindent
{\it Proof.} 
From the theorem hypothesis ($I \in L_S$) and from statement $(b)$ of theorem \ref{teosymm} ($\,\mathfrak{X}_i(R) = - {R}\langle \nabla | \mathfrak{X}_i \rangle \,$) it follows directly that the vector fields $\mathfrak{X}_i$ admit $R$ as an integrating factor. The SC result implies that the vector fields $\mathfrak{X}_i$ (for any $i \in \{1,2,3\}$) admit Darboux functions $R_i$ as Darboux integrating factors. So, we can write $R_i= {\cal F}_i(I)\,R\,$, where ${\cal F}_i(I)$ are functions of the first integral $I$. Therefore, the $R_i$ are also integrating factors for the vector field $\mathfrak{X}$. Since each $R_i$ is a Darboux integrating factor (in one of the pairs $(x,y)$, $(x,z)$ or $(y,z)$) and all are integrating factors for the vector field $\mathfrak{X}$, then $R_i= {\cal F}_{ij}(I)\,R_j$, where ${\cal F}_{ij}(I)$ are functions of the first integral $I$. This implies that either there is a Darboux first integral since ${\cal F}_{ij}(I)=\frac{R_i}{R_j}$ (at least one of the Jacobians $\frac{\partial(R_i,R_j)}{\partial(x_k,x_l)}$ is $\neq 0, i,\,j,\,k,\,l \in \{1,2,3\}, \, i \neq j, \, k \neq l)$, in which case there is (certainly) a Darboux integrating factor, or ${\cal F}_{ij}(I)=k_{ij}$ (where $k_{ij}$ are constants), i.e., $R_i= {k}_{ij}\,R_j$ implying that the $R_i$ are, in fact, Darboux functions on the three variables $(x,y,z)$ (the same function up to a multiplicative constant) and so, the vector field $\mathfrak{X}$ admits a Darboux integrating factor.$\,\,\,\Box$

\medskip

%%%%%%%%%%%%%%%%%%%%%%%%%%%%%%%%%%%%%
\subsection{The vector fields associated with $\mathfrak{X}_i$}
\label{bpvf}

In \cite{Nosjde2021} the authors proposed a new way of determining Darboux polynomials present in the integrating factor of plane polynomial vector fields: The basic idea was to construct another polynomial vector field such that its first integral is a Darboux integrating factor of the original vector field. In this section we will show how to adapt this idea to the current problem. 

%%%%%%%%%%%%%%%%
\subsubsection{Some important results}
\label{sir}

To begin with, we will present some results obtained in \cite{Nosjde2021} (unless by proposition \ref{x1pol}).

\medskip

\begin{defin}
\label{vfa}
Let $X_0 \equiv f_0 \,\partial_x + g_0 \,\partial_y$ ($f_0$ and $g_0$ are coprime polynomials in $\C[x,y]$) be a polynomial vector field presenting a Liouvillian first integral $I_0$ and, consequently, a Darboux integrating factor $R_0$. Let $X_1 \equiv f_1 \,\partial_x + g_1 \,\partial_y$ (where $f_1$ and $g_1$ are coprime polynomials in $\C[x,y]$) be another polynomial vector field such that $X_1(R_0)=0$. We call $X_1$ an {\bf associated vector field through the integrating factor} $R_0$. 
\end{defin}

\begin{obs}
\label{classx1}
Since any function of the first integral $I_0$ (which is invariant under the flow of the vector field $X_0$) multiplied by the integrating factor $R_0$ is itself an integrating factor $\overline{R}_0$ ($\overline{R}_0 \equiv {\cal G}(I_0)\,R_0$), $\overline{X}_1(\overline{R}_0)=0$ defines an equivalence class of vector fields: $\left[\overline{X}_1\right]$.
\end{obs}

\medskip

To prove the following theorem first we need to prove that there is always a polynomial vector field such that $X_1(R_0)=0$.

\begin{prop}
\label{x1pol}
Let $X_0$ be a polynomial vector field defined as above. Then, there exists a polynomial vector field $X_1$ such that $X_1(R_0)=0$.
\end{prop}

\medskip

\noindent
{\it Proof.} 
Since $R_0$ is a Darboux function, it can be written as $R_0 = {\rm e}^{Z_0}\,\prod_i {p_i}^{n_i}$, where $Z_0 \in \C(x,y)$ is a rational function, $p_i \in \C[x,y]$ are irreducible polynomials and $n_i$ are constants. This implies 
\begin{equation} 
\frac{\partial_i \left(R_0\right)}{R_0} = \partial_i \left( \ln(R_0)\right) =  \partial_i \!\left( Z_0+\sum_j n_j  \partial_i \frac{\left( p_i\right)}{p_i} \right), \,\,\, i\in \{1,2\}. \nonumber
\end{equation}
Since $Z_0$ is a rational function and $p_i$ are irreducible polynomials, we have that the terms $\,Z_0+\sum_j n_j  \partial_i \frac{\left( p_i\right)}{p_i}\,$ are rational functions of $(x,y)$ implying that $\,\partial_i \left( \ln(R_0)\right)$ are rational functions of $(x,y)$. Defining $\phi_1 \equiv -\partial_x \left( R_0\right)/\partial_y \left(R_0\right)$, we have that
\begin{equation} 
-\frac{\partial_x \left( \ln(R_0)\right)}{\partial_y \left( \ln(R_0)\right)} = -\frac{\partial_x \left(R_0\right)}{\partial_y \left(R_0\right)} = \phi_1.
\end{equation}
Therefore $\phi_1$ is a rational function of $(x,y)$. Defining the polynomials $f_1$ and $g_1$ as the  denominator  and numerator of $\phi_1$ (respectively), we have that the polynomial vector field $X_1 \equiv f_1\,\partial_x + g_1\,\partial_y$ has $R_0$ as a first integral, i.e., $X_1\left(R_0\right)=0.\,\,\,\Box$

\begin{teor}
Let $X_0$ and $X_1$ be polynomial vector fields defined as in definition {\em \ref{vfa}} above. Then 
\begin{equation}
\label{defcali}
g_0\,f_1-g_1\,f_0 = \frac{R_0}{R_1}\,\left({f_0}_x+{g_0}_y\right),
\end{equation}
where $R_1$ is an integrating factor for the vector field $X_1$.
\end{teor}

\noindent
{\bf Proof:} By hypothesis, we have that $X_1(R_0)=0$ (i.e., $R_0$ is a first integral of $X_1$). Therefore ${R_0}_x = R_1\,g_1, \, \, {R_0}_y = - R_1\,f_1$, implying that
$$
g_0\,f_1-g_1\,f_0 = g_0\,\frac{-{R_0}_y}{R_1} -  f_0\,\frac{{R_0}_x}{R_1} = - \frac{X_0(R_0)}{R_1} = \frac{R_0}{R_1} \, \langle \nabla | X_0 \rangle = \frac{R_0}{R_1} \, ({f_0}_x+{g_0}_y).\,\,\,\Box
$$

\begin{cor}
\label{coro1}
$\displaystyle{\frac{R_0}{R_1}}$ is an inverse integrating factor for the vector field $X_1$.
\end{cor}

\noindent
{\bf Proof:} Since $R_0$ is a first integral for the vector field $X_1$ then $\frac{R_1}{R_0}$ is also an integrating factor for the vector field $X_1.\,\,\,\Box$

\begin{cor}
\label{coro2}
$\displaystyle{\frac{R_0}{R_1}}$ is a polynomial or $\langle \nabla | X_0 \rangle$ has a polynomial factor in common with $R_1$ or $R_0$.
\end{cor}

\noindent
{\bf Proof:} Since $g_0\,f_1-g_1\,f_0$ is a polynomial, then $\displaystyle{\frac{R_0}{R_1}\, ({f_0}_x+{g_0}_y)}$ is a polynomial. So the conclusion follows directly.$\,\,\,\Box$

%\begin{obs}
%\label{r0sr1p}
%Note that the case in which $R_0/R_1$ is not a polynomial is easily treatable since $\langle \nabla | X_0 \rangle$ is a known polynomial (see section {\em \%ref{fincon}}, equation {\em (\ref{odesecper})}). For this reason, we can focus on the general case, i.e., where $R_0/R_1$ is a polynomial.
%\end{obs}

\begin{obs}
\label{mnplin}
If $R_0/R_1$ is a polynomial, we can write the equation {\em (\ref{defcali})} as 
\begin{equation}
\label{defcali2}
g_0\,f_1-g_1\,f_0 - \Upsilon \,\left({f_0}_x+{g_0}_y\right) = 0,   \,\,\,\,\, \Upsilon \equiv \frac{R_0}{R_1},
\end{equation}
and notice that it is linear in the unknown polynomials $f_1,\,g_1$ and $\Upsilon$.
\end{obs}

\medskip

%%%%%%%%%%%%%%%%
\subsubsection{Defining vector fields ${\cal X}_i$ such that ${\cal X}_i(R)=0$}
\label{dpvf}

In this subsection, we will apply the knowledge that was highlighted in remark \ref{mnplin}, together with the results shown in the previous subsection, to define three 2D polynomial vector fields ${\cal X}_i$ associated with $\mathfrak{X}_i$. First, remember that the three polynomial vector fields $\mathfrak{X}_i$ present the same Liouvillian first integral $I$ and the same Darboux integrating factor $R$ of $\mathfrak{X}$. Thus, there exist polynomial vector fields ${\cal X}_i$, which present $R$ as their first integral. 

\begin{defin}
\label{vfaxi}
Let $I \in L_S$, $R$, $\mathfrak{X}$ and $\mathfrak{X}_i$ defined as above. We define the polynomial vector fields ${\cal X}_i \equiv - \sum_{j,k} \epsilon_{ijk} \,{\cal P}_j \,\partial_k$  (${\cal P}_i$ are coprime polynomials in $\C[x,y,z]$) by $\,{\cal X}_i(R)=0.$
\end{defin}

\begin{obs}
\label{classxi}
As observed in remark {\em \ref{classx1}}, since any function of the first integral $I$ multiplied by the integrating factor $R$ is itself an integrating factor, the equation $\overline{{\cal X}}_i(R)=0$ defines three equivalence classes of vector fields: $[\overline{{\cal X}}_i],\,\,i \in \{1,2,3\}$.
\end{obs}

\begin{teor} 
\label{teosymm1}
Let $I \in L_S$, $R$, $\mathfrak{X}$, $\mathfrak{X}_i$ and ${\cal X}_i$ be defined as above. Then, the vector fields ${\cal X}_i$ present a Darboux integrating factor ${\cal R}$ such that $\Upsilon \equiv \frac{R}{{\cal R}}$ is an inverse integrating factor for them, i.e., $ {\cal X}_i (\Upsilon) = \Upsilon\,\langle \nabla | {\cal X}_i \rangle$. Besides, in the general case (i.e., the divergences $\langle \nabla | \mathfrak{X}_i \rangle$ do not have a common polynomial factor with $R$ or ${\cal R}$), $\Upsilon$ is a polynomial. 
\end{teor}

\noindent
{\it Proof.} The proof follows directly from the proofs of the corollaries \ref{coro1} and \ref{coro2}. 

\begin{teor} 
\label{teosymm2}
Let the polynomials ${\cal P}_i$, $P_i$ and $\Upsilon$ be as above. Then, 
\begin{equation}
\label{eqlinxcal}
\langle \partial_i | \mathfrak{X}_i \wedge {\cal X}_i \rangle = \Upsilon \,\langle \nabla | \mathfrak{X}_i \rangle.
\end{equation}
\end{teor}

\noindent
{\it Proof.} The proof follows directly from remark \ref{mnplin}. 

\begin{obs}
\label{obse1}
The equations $\langle \partial_i | \mathfrak{X}_i \wedge {\cal X}_i \rangle = \Upsilon \,\langle \nabla | \mathfrak{X}_i \rangle$ are linear in the unknown polynomials $\,{\cal P}_i$ and $\Upsilon$.
\end{obs}

%%%%%%%%%%%%%%%%%%%%%%%%%%%%%%%%%%%%%
\subsection{Constructing a linear probabilistic algorithm}
\label{lpa}

In this section we are going to `put the pieces together' and build a procedure to compute a Darboux integrating factor $R$ of 2ODE (\ref{r2ode}). We start by showing that the vector fields $\mathfrak{X}_i$ and ${\cal X}_i$ present a property that will be very important in this construction: since $\mathfrak{X}_i$ share the integrating factor $R$ with $\mathfrak{X}$, they also share the Darboux polynomials (present in the integrating factor $R$) with $\mathfrak{X}$ and ${\cal X}_i$. This implies that the polynomial $\Upsilon$ is, in general, formed by Darboux polynomials that are shared by the vector fields $\mathfrak{X}_i$ and ${\cal X}_i$. To prove this, let us first discuss some concepts and define some points more precisely.

\medskip

\noindent
Consider that the rational 2ODE (\ref{r2ode}) presents a first integral $I \in L_S$. We have seen that this implies that the associated vector field $\mathfrak{X}$ has a Darboux integrating factor $R$. Since any function of the first integral $I$ multiplied by an integrating factor is also an integrating factor, if the first integral $I$ is non-elementary, then $\mathfrak{X}$ has only one Darboux integrating factor (unless of a multiplicative constant) since, in this case, all other integrating factors (associated with the first integral $I$) will be non-elementary. So, in this case, there is a natural choice to define the integrating factor that represents the equivalence class $[ \overline{{\cal X}}_i ]$. In the case where the vector field $\mathfrak{X}$ presents a non-rational elementary first integral, we can apply the result of Prelle and Singer \cite{PreSin} ({\em For a plane polynomial vector field presenting an elementary first integral, there exists an integrating factor which is a $k^{th}$ root ($k \in \N$) of a rational function}) to the vector fields $\mathfrak{X}_i$ and choose an algebraic integrating factor to represent the class $[ \overline{{\cal X}}_i ]$. Finally, if the 2ODE presents a rational first integral, we will consider the representative $R$ of the class as the one with the minimum degree\footnote{As in this case the integrating factor will be rational, i.e., $R=num_R/den_R$, the degree of the integrating factor refers to the positive integer $\delta$ defined by $\delta = \max\{\deg(num_R),\deg(den_R)\}$.}.

\begin{defin}
\label{Ris}
Consider that the rational {\em 2ODE (\ref{r2ode})} presents a first integral $I \in L_S$. Then

\noindent
$(i)$ If $I$ is non-elementary, the integrating factor $R$ will be written as $R= {\rm e}^{A/\!B}\!\prod_j {p_j}^{n_j}$, where $p_j$, $A$ and $B$ are polynomials in $\C[x,y,z]$, $p_j$ are irreducible, $A$ and $B$ are coprime polynomials and $B$ is a Darboux polynomial (possibly not irreducible).

\medskip
\noindent
$(ii)$ If $I$ is elementary non-rational, the integrating factor $R$ will be written as $R= \prod_j {p_j}^{n_j}$ ($p_j$ in $\C[x,y,z]$ are irreducible polynomials and  ${n_j}$ are rational numbers).

\medskip
\noindent
$(iii)$ If $I$ is rational, the integrating factor $R$ will be written as $R= \prod_j {p_j}^{n_j}$ ($p_j$ in $\C[x,y,z]$ are irreducible polynomials and ${n_j}$ are integers such that $\delta \equiv \max\{\deg(num_R),\deg(den_R)\}$ is minimal).
\end{defin}

\begin{teor} 
\label{teopdarb}
Consider that the rational {\em 2ODE (\ref{r2ode})} presents a first integral $I \in L_S$. Let the polynomial vector fields $\mathfrak{X}_i$, ${\cal X}_i$, the Darboux integrating factors $R$ and ${\cal R}$ be defined as above and let $\Upsilon \equiv R/{\cal R}$ be a polynomial. Then, the following statements hold: 

\vspace{1mm}
\noindent
$(a)\,\,$ The Darboux polynomials $p_j$ of $\mathfrak{X}_i$ that are factors of $R$ or $B$, are also Darboux polynomials of ${\cal X}_i$. 

\vspace{2mm}
\noindent
$(b) \,\,$ The polynomial $\Upsilon$ has the following structure: $\Upsilon \!=\! \prod_j {p_j}^{k_j},\,k_j \in \N$.
\end{teor}

\bigskip

\noindent
{\it Proof.} 
\noindent
We will prove the theorem for the case where $I$ is a non-elementary Liouvillian function (item ($i$) of definition \ref{Ris}). The proof extends to the other itens: ($ii$) and ($iii$).

\medskip
\noindent
$(a) \,\,$ By definition we have that ${\cal X}_i(R)=0$. 
So, we can write:
\begin{equation}
\label{demoteodarb1}
\frac{{\cal X}_i(R)}{R} = {\cal X}_i\left(\frac{A}{B}\right) + \sum_j n_j \frac{{\cal X}_i(p_j)}{p_j} = 0.
\end{equation}
Multiplying both sides by $\prod_{j}p_j$, we can write:
\begin{equation} 
\label{demoteodarb2}
\prod_{j}p_j\,\, \frac{B\,{\cal X}_i(A) - A\,{\cal X}_i(B)}{B^2} + \sum_l c_l (\prod_{j,\,j\neq l}p_j) {\cal X}_i(p_l) = 0.
\end{equation}
Since the term $\sum_l c_l (\prod_{j,\,j\neq l}p_j) {\cal X}_i(p_l)$ is a polynomial, so is $\prod_{j}p_j\,\, \frac{B\,{\cal X}_i(A) - A\,{\cal X}_i(B)}{B^2}$. Since $B^2$ is a square, it cannot divide $\prod_{j}p_j$. Therefore, we have two possible situations:
\begin{itemize}
\item  None of the $p_j$ is a factor of $B$.
\item  Some of the $p_j$ are factors of $B$.
\end{itemize}

\begin{enumerate}
\item {\bf First situation:} ${\cal X}_i\left(\frac{A}{B}\right) = \frac{B\,{\cal X}_i(A) - A\,{\cal X}_i(B)}{B^2}$ is a polynomial. This implies that $ \sum_j n_j \frac{{\cal X}_i(p_j)}{p_j}$ is a polynomial $\Rightarrow p_j | {\cal X}_i(p_j)$. Therefore, multiplying  ${\cal X}_i\left(\frac{A}{B}\right)$ by $B$, we have that $B\,{\cal X}_i\left(\frac{A}{B}\right) = {\cal X}_i(A) - A\,\frac{{\cal X}_i(B)}{B}$ is a polynomial. Since $A$ and $B$ are coprime, $B\, | \,{\cal X}_i(B)$ implying that the irreducible Darboux polynomials of the vector fields $\mathfrak{X}_i$ that are factors of $B$ are also Darboux polynomials of ${\cal X}_i$.
\medskip

\item {\bf Second situation:} We will set up the following notation: $\!\!B \!=\! \beta \, \theta,\, \prod_{j}p_j \!= \! \Gamma \, \theta$, where $\theta$ is the commom factor. So, $\prod_{j}p_j\, \frac{B\,{\cal X}_i(A) - A\,{\cal X}_i(B)}{B^2} \!=\! \frac{\Gamma}{\beta} \left(\!{\cal X}_i(A) \!-\!  A\frac{{\cal X}_i(B)}{B}\right)$ is a polynomial. Multiplying by $\beta$ we obtain ${\Gamma}\,{\cal X}_i(A) \!-\! \Gamma A\frac{{\cal X}_i(B)}{B}$. Since $A$ and $\Gamma$ have no commom factors with $B$ then $B\, | \,{\cal X}_i(B)$ implying that $\theta\, | \,{\cal X}_i(\theta)$ and $\beta\, | \,{\cal X}_i(\beta)$. It remains to prove that the polynomials $p_j$ that are not factors of $B$ (i.e., the factors of $\Gamma$) are also Darboux polynomials of ${\cal X}_i$. We have that
\begin{equation}
\label{demoteodarb3}
\frac{{\cal X}_i(R)}{R} = \frac{{\cal X}_i\left({\rm e}^{A/B}\,\prod_j {p_j}^{n_j}\right)}{{\rm e}^{A/B}\,\prod_j {p_j}^{n_j}}= {\cal X}_i\left(\frac{A}{B}\right) + \frac{{\cal X}_i(\prod_j {p_j}^{n_j})}{\prod_j {p_j}^{n_j}} = 0.
\end{equation}
Since the polynomials that are factors of $\theta$ are Darboux polynomials of ${\cal X}_i$, then 
\begin{equation}
\label{demoteodarb4}
\Gamma\,\frac{{\cal X}_i(\prod_j {p_j}^{n_j})}{\prod_j {p_j}^{n_j}} 
\end{equation}
is a polynomial. So, we have that $\Gamma\,{\cal X}_i\left(\frac{A}{B}\right) = \Gamma \frac{B\,{\cal X}_i(A) - A\,{\cal X}_i(B)}{B^2}$ is a polynomial. Since, $\Gamma$ and $B^2$ have no factors in common, then ${\cal X}_i\left(\frac{A}{B}\right)$ is a polynomial implying that $\frac{{\cal X}_i(\prod_j {p_j}^{n_j})}{\prod_j {p_j}^{n_j}}$ is a polynomial and, therefore, $p_j | {\cal X}_i(p_j)$. 
\end{enumerate}
This proves the first statement.
\medskip

\noindent
$(b) \,\,$ By definition $\Upsilon = R/{\cal R}$ and $\partial_i(R) = {\cal R}\,{\cal P}_i$. So:
\begin{eqnarray} 
R_x \!\!\!\!&=&\!\!\!\! \left(\partial_x\left(\frac{A}{B}\right)+\frac{\sum_k n_k\,\partial_x(p_k)\,\prod_{l,l\neq k}p_l}{\prod_j p_j}\right)\,{\rm e}^{A/B}\,\prod_j {p_j}^{n_j}  \nonumber \\ [2mm]
\!\!\!\!&=&\!\!\!\! \frac{(A_xB-B_xA){\prod_j p_j}+B^2\sum_k n_k\,\partial_x(p_k)\,\prod_{l,l\neq k}p_l }{B^2\prod_j p_j}\,R = \frac{Pol_{[x]}\,R}{B^2\prod_j p_j}= {\cal P}_1\,{\cal R}, \nonumber 
\end{eqnarray}

\begin{eqnarray}
R_y\!\!\!\!&=&\!\!\!\! \frac{(A_yB-B_yA){\prod_j p_j}+B^2\sum_k n_k\,\partial_y(p_k)\,\prod_{l,l\neq k}p_l }{B^2\prod_j p_j}\,R = \frac{Pol_{[y]}\,R}{B^2\prod_j p_j}= {\cal P}_2\,{\cal R}, \nonumber \\ [2mm]
R_z\!\!\!\!&=&\!\!\!\! \frac{(A_zB-B_zA){\prod_j p_j}+B^2\sum_k n_k\,\partial_z(p_k)\,\prod_{l,l\neq k}p_l }{B^2\prod_j p_j}\,R= \frac{Pol_{[z]}\,R}{B^2\prod_j p_j}= {\cal P}_3\,{\cal R}, \nonumber \\ [2mm]
\!\!\!\!&\Rightarrow&\!\!\! \Upsilon = \frac{R}{\cal R} = \frac{{\cal P}_1}{Pol_{[x]}}\,B^2\prod_j p_j = \frac{{\cal P}_2}{Pol_{[y]}}\,B^2\prod_j p_j = \frac{{\cal P}_3}{Pol_{[z]}}\,B^2\prod_j p_j.
\end{eqnarray}
This implies $\frac{{\cal P}_1}{Pol_{[x]}} = \frac{{\cal P}_2}{Pol_{[y]}} = \frac{{\cal P}_3}{Pol_{[z]}}$. Since ${\cal P}_i$ are coprime and $\Upsilon$ is polynomial, then $\frac{Pol_{[x]}}{{\cal P}_1} = \frac{Pol_{[y]}}{{\cal P}_2} = \frac{Pol_{[z]}}{{\cal P}_3} = \rho$, where $\rho$ is a polynomial (or a constant) such that $\rho\,|\,B^2\prod_j p_j\,$. Since $\,\Upsilon = \frac{B^2\prod_j p_j}{\rho}$ and the term $\,B^2\prod_j p_j\,$ is formed by products of irreducible Darboux polynomials, then $\,\Upsilon = \prod_u {p_u}^{k_u},\,\,k_u \in \N$. This proves the second statement. $\,\,\,\Box$

\medskip

\noindent
In what follows, we will make use of the statements of theorem \ref{teopdarb} to construct a probabilistic linear algorithm ({\it PLDIF}):

\medskip
\noindent
{\bf Procedure {\it PLDIF} (sketch):}

\begin{enumerate}

\item Construct the operators $\mathfrak{X},\,\mathfrak{X}_1,\,\mathfrak{X}_2,\,\mathfrak{X}_3$.

\item Construct four polynomials ${{\cal P}_1}_c,\,{{\cal P}_2}_c,\,{{\cal P}_3}_c, \, \Upsilon_c$ of degrees $d_1,\,d_2,\,d_3$ and $d_{\Upsilon}$, respectively, with undetermined coefficients. 

\item Substitute then in one of the equations (\ref{eqlinxcal}):
\begin{eqnarray}
E_1 &:=& {\cal P}_2\,{P_3} - {P_2}\,{\cal P}_3 = \Upsilon \,\langle \nabla | \mathfrak{X}_1 \rangle, \label{e1} \\
E_2 &:=& {\cal P}_1\,{P_3} - {P_1}\,{\cal P}_3 = \Upsilon \,\langle \nabla | \mathfrak{X}_2 \rangle, \label{e2} \\
E_3 &:=& {\cal P}_1\,{P_2} - {P_1}\,{\cal P}_2 = \Upsilon \,\langle \nabla | \mathfrak{X}_3 \rangle. \label{e3}
\end{eqnarray}

\item  Collect the chosen equation $E_i$ in the variables $(x,y,z)$ obtaining a set of linear equations $S_{E_i}$ for the coefficients of the polynomial candidates. 

\item Solve $S_{E_i}$ to the undetermined coefficients, obtaining a solution $Sol_{sys}$. 

\item Replace the solution $Sol_{sys}$ in the candidate ${\Upsilon}_c$, collect it with respect to the remaining coefficients and factor the polynomials that multiply each one of them (the remaining coefficients).

\item Select which of these polynomial factors are Darboux polynomials of the vector field $\mathfrak{X}$.

\item Rebuild the candidate ${\Upsilon}_c$ by adding the Darboux polynomials found to it (for example, if a Darboux polynomial $p_1$ of degree $d_1$ was found, reconstruct the candidate using $\Upsilon_c = \Upsilon_p \,p_1$, where the degree of $\Upsilon_p$ is $d_u - d_1$).

\item Redo the steps $3 \rightarrow 8$ until no new Darboux polynomials appear 

\item If all the indeterminates have been found, substitute the complete solution into the polynomial candidates, obtaining ${\cal P}_j,\,{\cal P}_k,\,\Upsilon$. If not, increase the degree of candidates and redo the steps $2 \rightarrow 9$. 

\item From  ${\cal P}_j,\,{\cal P}_k,\,\Upsilon$ determine $R$ by quadratures.

\item Use $R$ to compute the first integral $I$ of the rational 2ODE by quadratures.
\end{enumerate}

\vspace{1mm}
\begin{obs} \label{keyp}
The key points for the efficiency of the algorithm are:
\begin{itemize}
\item The equations $E_i$ ($\langle \partial_i | \mathfrak{X}_i \wedge {\cal X}_i \rangle = \Upsilon \,\langle \nabla | \mathfrak{X}_i \rangle$) are linear in the coefficients of the unknown polynomials ${\cal P}_1,\,{\cal P}_2,\,{\cal P}_3,\,\Upsilon$.
\item There are three of them.
\item $\Upsilon$ is a polynomial formed by Darboux polynomials present in the integrating factor $R$.
\item $\Upsilon$ is an inverse integrating factor for the vector fields ${\cal X}_i$.
\end{itemize}
\end{obs}

\medskip

\begin{exem} \label{eximp1} \end{exem}Consider the following 2ODE:
\begin{eqnarray}
z' &=& z \left( {x}^{7}y{z}^{2}-{x}^{6}{z}^{3}-2\,{x}^{5}{y}^{2}z+{x}
^{5}y{z}^{2}+2\,{x}^{4}y{z}^{2}-{x}^{4}{z}^{3}-{x}^{5}z-2\,{x}^{4}{y}^
{2}+{x}^{3}{y}^{3} \right. \nonumber \\ [2mm]
&& \left. -2\,{x}^{3}{y}^{2}z+{x}^{3}{y}^{2}+4\,{x}^{3}yz-{x}^
{2}{y}^{2}z+2\,{x}^{2}y{z}^{2}+{x}^{3}y-{x}^{3}z-2\,{x}^{2}yz-2\,{x}^{
2}{z}^{2}  \right. \nonumber \\ [2mm]
&& \left. +x{y}^{3}+x{z}^{2}-{y}^{2}z+yx \right) / \left( {x}^{5}y{z}^
{2}-{x}^{4}{z}^{3}+{x}^{4}{y}^{2}-2\,{x}^{3}{y}^{2}z-2\,{x}^{3}yz   \right. \nonumber \\ [2mm]
&& \left. +2\,{x}^{2}y{z}^{2}-{x}^{3}z+{x}^{2}{z}^{2}+x{y}^{3}-{y}^{2}z+yx \right) x. \label{eqeximp1}
\end{eqnarray}

\begin{itemize}
\item The procedure {\it NLS} finds the symmetry (and therefore the vector fields $\mathfrak{X}_i$)  in 0.03 seconds: 
\begin{eqnarray}
\!\!\!P_1\!\!\!\! &=&\!\!\!\!\! -z \!\left(\! {x}^{5}\!y{z}^{2}\!-\!{x}^{4}\!{z}^{3}\!-\!2{x}^{4}\!{y}^{2}\!-\!2{x}^{3}\!{y}^{2}\!z\!+\!4{x}^{3}\!yz\!+
\!2{x}^{2}\!y{z}^{2}\!-\!{x}^{3}\!z\!-\!2{x}^{2}\!{z}^{2}\!+\!x{y}^{3}\!-\!{y}^{2}\!z\!+\!yx \!\right),
 \nonumber \\ [2mm]
\!\!\!P_2\!\!\!\! &=&\!\!\!\!\! - \!\left(\! {x}^{6}\!y{z}^{2}\!-\!{x}^{5}\!{z}^{3}\!-\!2{x}^{4}\!{y}^{2}\!z\!+\!2{x}^{3}\!y
{z}^{2}\!-\!{x}^{4}\!z\!+\!{x}^{2}\!{y}^{3}\!+\!{x}^{2}\!{y}^{2}\!-\!x{y}^{2}\!z\!+\!{x}^{2}\!y\!-\!2xyz\!+\!{z}^{2}\! \right)\! x, \nonumber \\ [2mm]
\!\!\!P_3\!\!\!\! &=&\!\!\!\!\! \left(\! {x}^{5}\!y{z}^{2}\!-\!{x}^{4}\!{z}^{3}\!+\!{x}^{4}\!{y}^{2}\!-\!2{x}^{3}\!{y}^{2
}\!z\!-\!2{x}^{3}\!yz\!+\!2{x}^{2}\!y{z}^{2}\!-\!{x}^{3}\!z\!+\!{x}^{2}\!{z}^{2}\!+\!x{y}^{3}\!-\!{y}^{2}\!z\!+\!yx\! \right)\! x.
\end{eqnarray}
\item The procedure {\it PLDIF}:
\begin{itemize}
\item Steps 1 and 2: $d_1=7,\,d_2=7,\,d_3=7$ and $d_{\Upsilon}=8$.

\item Steps 3, 4, 5 and 6: (the chosen equation was $E_3$) 405 unknown coefficients were reduced to only 4 (in 0.14 seconds). The candidates ${{\cal P}_1}_c,\,{{\cal P}_2}_c$ and $\Upsilon_c$, depending on these 4 coefficients still to be determined, can be expressed by:
\begin{eqnarray}
\!\!\!\!\!\!\!\!{{\cal P}_1}_c\!\!\!\!\! &=&\!\!\!\!\!-\frac{1}{2}\!\left( 7\,{x}^{4}{y}^{2}z-\!12\,{x}^{3}y{z}^{2}+{x}^{3}{z}^{2}-3\,
{x}^{2}{y}^{3}+5\,{x}^{2}{z}^{3}+4\,x{y}^{2}z-xyz-y{z}^{2} \right)\! a_{29} \nonumber \\ [2mm]
&& 
\!\!\!\!\!+\frac{1}{4}\! \left(14{x}^{3}\!{y}^{3}\!z\!+\!7{x}^{3}\!{y}^{2}\!z\!-\!22{x}^{2}\!{y}^{2}\!{z}^{2}\!-\!10{x}^{2}\!y{z}^{2}\!-\!6\,x{y}^{4}\!+\!8xy{z}^{3}\!+\!{x}^{2}\!{z}^{2}\!+\!5x{z}^{3}\!+\!6{y}^{3}\!z \right.  \nonumber \\ [2mm]
&& \left. 
\!\!\!\!\!\!-\!2{y}^{2}\!z\!-\!2yz \right)\!c_{120}+\!\left( 2{x}^{3}\!y{z}^{2}\!-\!2{x}^{2}\!{z}^{3}\!+\!3{x}^{2}\!{y}^{2}\!-
\!2x{y}^{2}\!z\!-\!4xyz\!+\!2y{z}^{2}\!-\!xz\!+\!{z}^{2}\! \right)\! c_{86} \nonumber \\ [2mm]
&&
\!\!\!\!\!-\frac{1}{2}\!\left( 14\,{x}^{4}\!y{z}^{2}\!+\!14{x}^{3}\!y{z}^{3}\!-\!16{x}^{3}\!{z}^{3}\!-\!14{x}^{2}\!{y}^{3}\!z\!-
\!8{x}^{2}\!{z}^{4}\!+\!15{x}^{2}\!{y}^{2}\!z\!+\!2x{y}^{2}\!{z}^{2}\!-\!2{x}^{2}\!yz \right.  \nonumber \\ [2mm]
&& \left. 
\!\!\!\!\!+6{y}^{4}\!-\!10{x}^{2}\!z-7\,x{z}^{2}-12\,{y}^{3}-{z}^{3}+6\,{y}^{2}+6\,y \right) c_{\,94},  \nonumber
\end{eqnarray}
\begin{eqnarray}
\!\!\!\!\!\!\!\!{{\cal P}_2}_c\!\!\!\! &=&\!\!\!\!\!-\frac{1}{2}x\!\left(\! 2\,{x}^{4}yz+{x}^{4}z-2\,{x}^{3}{z}^{2}-4\,{x}^{2}{y}^{
2}-{x}^{2}y+6\,xyz-2\,{z}^{2} \right)\!a_{29}\!+\frac{1}{4} \left( 6\,{x}^{4}{y}^{2}z \right. \nonumber \\ [2mm]
&& \left.
\!\!\!+4{x}^{4}\!yz\!-\!6{x}^{3}\!y{z}^{2}\!+\!{x}^{4}\!z\!-\!2{x}^{3}\!{z}^{2}\!-\!10{x}^{2}\!{y}^{3}\!-\!4{x}^{2}\!{y}^{2}\!
+\!14x{y}^{2}\!z\!-\!2{x}^{2}\!y\!+\!4xyz\!-\!4y{z}^{2}\right. \nonumber \\ [2mm]
&& \left.
\!-2{z}^{2} \right)c_{120}+{x}^{2} \!\left( 2{x}^{3}\!yz-\!2{x}^{2}\!{z}^{2}\!-\!2x{y}^{2}\!+2xy+\!2yz\!-\!x\!-
\!2z \right) c_{86} +\frac{1}{2}\!\left(2{x}^{5}\!{z}^{2} \right. \nonumber \\ [2mm]
&&
\!-6{x}^{4}{z}^{3}\!+6{x}^{3}{y}^{2}\!z-8{x}^{3}yz+10{x}^{2}y{z}^{2}\!-\!{x}^{3}\!z\!-\!6{x}^{2}\!{z}^{2}\!-
10x{y}^{3}\!-4x{z}^{3}\!+\!18x{y}^{2}  \nonumber \\ [2mm]
&& \left. 
\!+4{y}^{2}\!z+2xy\!-\!6yz\!-\!2x\!-\!4z \right) c_{94},
\end{eqnarray}
\begin{eqnarray}
\!\!\!\!\!\!\!\Upsilon_c\!\!\! &=&\!\!\!-\frac{1}{2} \,x \left( xy-z \right) ^{2} \left( {x}^{2}z-y \right) a_{29}+
\frac{1}{4}\left( 2{x}^{2}yz+{x}^{2}z-\!2{y}^{2} \right) \left( xy-z \right) ^{2}c_{\,120}  \nonumber \\ [2mm]
&&
\!\!\!+ x \left( xy-z \right) ^{2} c_{\,86}\,-\,\frac{1}{2}\left( 2{x}^{5}\!y{z}^{2}\!+\!2{x}^{4}\!y{z}^{3}\!-\!2{x}^{4}\!{z}^{3}\!-\!2{x}
^{3}\!{y}^{3}\!z\!-\!2{x}^{3}\!{z}^{4}\!+\!{x}^{3}\!{y}^{2}\!z\! \right. \nonumber \\ [2mm]
&& \left. 
\!\!\!\!-2{x}^{3}\!yz\!+\!2x{y}^{4}\!+\!2xy{z}^{3}\!-\!2{x}^{3}\!z-\!4x{y}^{3}\!-\!x{z}^{3}\!-2{y}^{3}\!z\!+\!2{y}^{2}\!x\!+\!4{y}^{2}\!z\!+
\!2xy \right) c_{\,94}. \nonumber
\end{eqnarray}

\item Step 7: the Darboux polynomials found are $\{\,x, \,(xy-z)^2,\,\,x^2z-y\}$ (without spending any measurable computational time).

\item Steps 8,9 and 10: 
\begin{eqnarray}
{\cal P}_1 &=& 7\,{x}^{4}{y}^{2}z-12\,{x}^{3}y{z}^{2}+{x}^{3}{z}^{2}-3\,{x}^{2}{y}^{3
}+5\,{x}^{2}{z}^{3}+4\,x{y}^{2}z-xyz-y{z}^{2}, \nonumber \\ [2mm]
{\cal P}_2 &=& x \left( 2\,{x}^{4}yz+{x}^{4}z-2\,{x}^{3}{z}^{2}-4\,{x}^{2}{y}^{2}-{x}
^{2}y+6\,xyz-2\,{z}^{2} \right), \nonumber \\ [2mm]
\Upsilon &=& \left( {x}^{2}z-y \right) \left( xy-z \right) ^{2} x. \nonumber 
\end{eqnarray}

\item Steps 11 and 12: Since ${\cal R}=1/\Upsilon$ is an integrating factor for the vector field ${\cal X}_3 = {\cal P}_2\partial_x-{\cal P}_1\partial_y$, we can determine the first integral of ${\cal X}_3$ (i.e., the integrating factor $R$ of the vector field $\mathfrak{X}$) with simple quadratures:
\begin{equation}
R = \frac{{\rm e}^{\frac {x}{xy-z}}}{\left( {x}^{2}z-y \right) \left( xy-z \right) ^{2} x}.
\end{equation}
Finally, with the integrating factor $R$, we can obtain the Liouvillian first integral of the vector field $\mathfrak{X}$:
\begin{equation}
I = \frac{{\rm e}^{{\frac {x}{xy-z}}}}{ \left( {x}^{2}z-y \right) }+{\it Ei}
 \left( 1,-{\frac {x}{xy-z}} \right).
\end{equation}
\end{itemize}

\end{itemize}

\medskip

\begin{obs}
\label{obs2}
Some comments:
\begin{enumerate}
\item In the example {\em \ref{eximp1}} above we get a massive determination of coefficients: a reduction from 405 undetermined coefficients to only 4. It is important to emphasize that this huge reduction is not an isolated case, as it happens in the vast majority of cases. 
\item Substituting the solution found (related to the item above) in candidate $\Upsilon_c$ and grouping the result into the remaining indeterminates, we obtain, after factoring each term of the sum, {\bf all} Darboux polynomials present in $\Upsilon$. 
\item In all tested cases we found all Darboux polynomials needed to build $\Upsilon$ (In the example {\em \ref{eximp1}} above, the polynomial that multiplies the coefficient $a_{29}$ is $\Upsilon$ itself).
\item The number of coefficients that remain undetermined (after the reduction process mentioned in the first part of the {\it PLDIF} procedure) can vary depending on which equation we use: $E_1$, $E_2$ or $E_3$. For instance, in the example just presented, if we use the equation $E_1$ instead of equation $E_3$, the reduction would be from 405 to 14. The number of Darboux polynomials found can also vary. However, in all cases studied, it was always possible to determine all of them with the strategy mentioned in the second item above, i.e., {\bf in all tested cases the problem of determining the Darboux polynomials was solved linearly in its entirety}.
\item If we start with lower degrees for the polynomials ${\cal P}_1,\,{\cal P}_2,\,{\cal P}_3,\,\Upsilon$ it is possible (in fact, quite common) to find Darboux polynomials in the middle of the process, that is, before reaching the required degree. In this way, the iteration process becomes much more efficient in the most complicated cases, i.e., in situations where the iteration (if we start from higher degrees for the candidates) does not result in new Darboux polynomials.
\item If the rational {\em 2ODE} presents an elementary first integral, the steps 10, 11 and 12 are not necessary because in this case we can directly find the multiplicity of the Darboux polynomials that form the integrating factor $R$ by using the final (linear) part of the Prelle-Singer method ( $\sum_j {{n_i}_j} {{q_i}_j}+\langle \nabla | \mathfrak{X}_i \rangle=0$, see {\em \cite{PreSin}} ).
\item Although the `most important' part of the {\it PLDIF} procedure is the determination of the Darboux polynomials, it is worth mentioning that, if $I$ is a non-elementary first integral (i.e., in which case the integrating factor $R$ will not be (in general) an algebraic function), the third part of the procedure (determining the multiplicity of the Darboux polynomials which are factors of $\Upsilon$) makes the process much more efficient. This is because, in this case, even having all the Darboux polynomials necessary for the construction of the integrating factor, we have no idea how they appear in the exponential factor, and this verification (testing all possible combinations) would exponentially increase the CPU time spent by the algorithm (see the comments by Guillaume Chèze in {\em \cite{Che,ChCo}} about the algorithm developed in {\em \cite{Nosjcam2005}} and implemented in {\em \cite{Noscpc2007}}). However, by determining the multiplicity of the Darboux polynomials that are factors of $\Upsilon$, we can use it (since $\Upsilon$ is an inverse integrating factor for the vector fields ${\cal X}_i$) to obtain the integrating factor $R$ of the {\em 2ODE} by quadratures.
\item In the vast majority of cases, one of the equations $E_1$, $E_2$, $E_3$ is enough to completely determine the polynomial $\Upsilon$ and, consequently, the polynomials ${\cal P}_1,\,{\cal P}_2,\,{\cal P}_3$ (as happened in the example we just showed).
\item The rational {\em 2ODE} {\em (\ref{eqeximp1})} (example {\em \ref{eximp1}}) presents a Liouvillian first integral that is not determined by the methods implemented in the solver ({\tt dsolve}) of the Maple (v.17) platform of symbolic computing\footnote{Maple is a general-purpose computer algebra system (i.e., a symbolic computing environment) which is also a multi-paradigm programming language. It can manipulate mathematical expressions and find symbolic solutions to ordinary and partial differential equations ({\em ODEs} and {\em PDEs}).}.
\item As we do not have an upper bound for the degree of the polynomials ${\cal P}_1,\,{\cal P}_2,\,{\cal P}_3$ and ${\Upsilon}$, the procedure {\it PLDIF} may not end (see step 10 of the procedure), and so {\it PLDIF} is actually a semi-algorithm. 
\item The degree of polynomials ${\cal P}_1,\,{\cal P}_2,\,{\cal P}_3$ and ${\Upsilon}$ is not determined (a priori) by the steps of the method (see later comments in section \ref{perfor}). Thus, it is not well determined which is the best (i.e., most efficient) iteration process for the degree of polynomial candidates.
It is also not clear if a given equation ($E_1$, $E_2$ or $E_3$) should be used first. For instance, in the example {\em \ref{eximp1}} shown above, it is much more efficient to use the equation $E_3$.

\end{enumerate}

\end{obs}

%%%%%%%%%%%%%%%%%%%%%%%%%%%%%%%%%%%%%%%%%%%%%%%%%%%%%%%%%%%%%%%%%%%%%
\section{Performance of the algorithms}
\label{perfor}

In this section we make a preliminary study of the performance of the {\it NLS} and {\it PLDIF} algorithms and some considerations about the theoretical questions still unanswered as well as possible improvements and extensions of the developed algorithms.

\subsection{Some `difficult' 2ODEs}
\label{exmps}

In this subsection we make a brief analysis of the performance of the constructed algorithms: we compare the efficiency of the {\it NLS} and {\it PLDIF} procedures (in a Maple pre-implementation) with the performance of the S-function method (see \cite{Noscpc2019}). For this, we build a small set of seven rational 2ODEs (presenting a Liouvillian first integral) which are divided into two subsets:
\begin{itemize}
\item In the first one, we build four 2ODEs, three of them with a non-elementary Liouvillian first integral, according to the following criteria: they are not solved by canonical procedures (implemented in Maple CAS); they do not have point symmetries and the $\lambda$-symmetries are very complex; the integrating factor is formed by Darboux polynomials of relatively high degree. The S-function method is considered to fail if it does not return the answer within 30 seconds or until a memory consumption of 300 MB is reached. Even if the symmetry is found (after applying the algorithm {\it NLS}), the associated 1ODE can not be solved by the methods implemented in Maple CAS (with a powerful ODE solver, the {\tt dsolve} command).
\item The second subset presents three 2ODEs that the S-function method can solve (or at least find the S-function) with more reasonable use of time/memory. 
\end{itemize}

\medskip

\subsubsection{First set:}

\begin{obs}
The following tables describe the memory and CPU time expenditures of the most costly routines (computationally speaking) of the procedures {\it NLS} and {\it PLDIF} when applied to the 2ODEs that follow.
\end{obs}

\noindent {\bf 2ODE-1:}
\begin{eqnarray}
z' &=& \!\!\left(2{x}^{5}{y}^{4}{z}^{2}\!-\!{x}^{5}{y}^{4}\!-\!2{x}^{5}{y}^{3
}z\!+\!4{x}^{3}{y}^{2}{z}^{4}\!-\!{x}^{4}{y}^{4}\!-\!4{x}^{4}{y}^{2}{z}^{2}\!+\!2{x}^{5}yz\!-\!2{x}^{3}{y}^{3}z \right.
\nonumber \\ [2mm]
&& 
\left. -\!2{x}^{3}{y}^{2}{z}^{2}-2{x}^{3}y{z}
^{3}\!+\!2x{z}^{6}\!+\!{x}^{4}{y}^{2}\!+\!2xy{z}^{3}\!-\!x{z}^{4}\!-\!{x}^{2}{z}^{2}\!+\!{z}^{4}\right)/
\nonumber \\ [2mm]
&& 
\left( -2x \left( {x}^{4}{y}^{5}+2\,{x}^{2}{y}^{3}{z}^{2}-{x}^{2}{y}^{2}
z+y{z}^{4}-2\,xy{z}^{2}+{x}^{2}z-{z}^{3} \right) \right) 
\label{s4ex1}
\end{eqnarray}

\begin{table}[h]
{\begin{center} {\footnotesize
\begin{tabular}
{|c|c|c|c|}
\hline
Algorithm & Task & Memory (MB) & Time (sec)  \\
\hline
 {\it NLS} &  $\{P_3,P_2,P_1\}$ & 1 & 0.047    \\
\hline
 {\it PLDIF} (part 1) & Coeff. reduction & 2 & 0.078   \\
\hline
 {\it PLDIF} (part 2) & Compute DPs & 0 & 0.032   \\
\hline
 {\it PLDIF} (part 3) & Compute $\Upsilon$ & 1 & 0.390   \\
\hline
 {\it NLSPLDIF} & Find a LFI $I$ & 4 & 0.547   \\
\hline
\end{tabular} }
\caption{Time and Memory consumption --- 2ODE (\ref{s4ex1})}
\end{center}}
\label{tabex1}
\end{table}

\noindent
{\bf Procedure {\it NLS}:} 

\noindent 1) {\it NLS} computes $P_2 \!=\! 2x\left( {x}^{4}{y}^{4}z\!-\!{x}^{4}{y}^{3}\!+\!2{x}^{2}{y}^{2}{z}^{3}\!-\!2
{x}^{3}{y}^{2}z+\!{x}^{4}y\!-\!{x}^{2}y{z}^{2}\!+\!{z}^{5} \right)$ and $P_1=-{x}^{5}{y}^{4}-{x}^{4}{y}^{4}-2\,{x}^{3}{y}^{3}z-2\,{x}^{3}{y}^{2}{z}^{2}+{x}^{4}{y}^{2}+2\,xy{z}^{3}-x{z}^{4}-{x}^{2}{z}^{2}+{z}^{4}$.

\medskip
\noindent
{\bf Procedure {\it PLDIF}:} 

\noindent 2) Reduction of undetermined coefficients (part 1): 469 $\rightarrow$ 14

\noindent 3) DPs found (part 2): $\{ p_1=x, p_2=y, p_3=z, p_4=x^2y^2+z^2\}$.

\noindent 4) Exponents found (part 3): $\{ n_1=1, n_2=0, n_3=0, n_4=2\}$. So, ${\cal P}_3=-2xz \left( 2{x}^{2}{y}^{2}\!+\!2{z}^{2}\!+x \right)$, ${\cal P}_2= -2{x}^{3}y\left( 2{x}^{2}{y}^{2}\!+\!2{z}^{2}\!+x \right)$, ${\cal P}_1=-5{x}^{4}{y}^{4}\!-\!6{x}^{2}{y}^{2}{z}^{2}\!-\!{x}^{3}{y}^{2}\!-\!{z}^{4}\!+\!x{z}^{2}\,$ and $\,\Upsilon= x\left( {x}^{2}{y}^{2}+{z}^{2} \right) ^{2}$.

\medskip

\begin{obs}
The other parts have a very small algorithmic cost compared to those shown in Table 1:

\noindent Since $\Upsilon$ is an inverse integrating factor for ${\cal X}_i$, we can find $R$ with simple quadratures: ${\cal R}_z=\frac{{\cal P}_3}{\Upsilon}$, ${\cal R}_y=\frac{{\cal P}_2}{\Upsilon}$, ${\cal R}_x=\frac{{\cal P}_1}{\Upsilon}$ and so
\begin{equation}
\label{s4ex1R}
R = \frac{{\rm e}^{{\frac {x}{{x}^{2}{y}^{2}+{z}^{2}}}}}{\left( {x}^{2}{y}^{2}+
{z}^{2} \right)x}.
\end{equation}
Therefore, since $I_z=R\,P_3$, $I_y=R\,P_2$, $I_x=R\,P_1$, we have
\begin{equation}
\label{s4ex1I}
I ={{\rm e}^{{\frac {x}{{x}^{2}{y}^{2}+{z}^{2}}}}} \left( -2\,yz+x
 \right) +{\it Ei} \left( 1,-{\frac {x}{{x}^{2}{y}^{2}+{z}^{2}}} \right).
\end{equation}
\end{obs}

\noindent {\bf 2ODE-2:}
\begin{eqnarray}
z' \!\!\!&=& \!\!\!\left({x}^{9}{z}^{5}\!+\!4{x}^{9}{z}^{4}\!-\!{x}^{8}y{z}^{4}\!-\!3{x}^
{6}{y}^{2}{z}^{3}\!-\!2{x}^{5}{y}^{3}{z}^{3}\!-\!4{x}^{5}{y}^{3}{z}^{2}\!+\!2{x}^{4}{y}^{4}{z}^{2}\!-\!
2{x}^{5}{z}^{3}\!+\!3{x}^{2}{y}^{5}z \right.
\nonumber \\ [2mm]
&& 
\left. \!+x{y}^{6}z\right)/\left(-2{x}^{5}z \left( {x}^{5}{z}^{2}-x{y}^{3}-x+y \right) \right).
\label{s4ex2}
\end{eqnarray}

\begin{table}[h]
{\begin{center} {\footnotesize
\begin{tabular}
{|c|c|c|c|}
\hline
Algorithm & Task & Memory (MB) & Time (sec)  \\
\hline
 {\it NLS} & $\{P_3,P_2,P_1\}$ & 1 & 0.047    \\
\hline
 {\it PLDIF} (part 1) & Coeff. reduction & 6 & 0.422   \\
\hline
 {\it PLDIF} (part 2) & Compute DPs & 0 & 0.141   \\
\hline
 {\it PLDIF} (part 3) & Compute $\Upsilon$ & 1 & 0.125   \\
\hline
 {\it NLSPLDIF} & Find a LFI $I$ & 8 & 0.735   \\
\hline
\end{tabular} }
\caption{Time and Memory consumption --- 2ODE (\ref{s4ex2})}
\end{center}}
\label{tabex2}
\end{table}

\noindent
{\bf Procedure {\it NLS}:} 

\noindent 1)$\!\!$ {\it NLS}$_{2I}\!$ computes$\!$ $P_2 \!\!=\!\! \left( {x}^{8}{z}^{4}\!-\!3{x}^{5}{y}^{2}{z}^{2}\!-\!2{x}^{4}{y}^{3}{z}^{2}\!-\!2{x}^{4}z^{2}\!+\!3x{y}^{5}\!+\!{y}^{6}\!+\!3{y}^{2}x\!-\!{y}^{3}\!+\!\!1\! \right)\! x$ and $P_1=4\,{x}^{9}{z}^{4}-{x}^{8}y{z}^{4}-4\,{x}^{5}{y}^{3}{z}^{2}+2\,{x}^{4}{y}^{4}{z}^{2}-4\,{x}^{5}{z}^{2}+6\,{x}^{4}{z}^{2}y-{y}^{7}-2\,{y}^{4}-y$.

\medskip
\noindent
{\bf Procedure {\it PLDIF}:} 

\noindent 2) Reduction of undetermined coefficients (part 1): 1235 $\rightarrow$ 1

\noindent 3) DPs found (part 2): $\{ p_1=x, p_2=y, p_3=z, p_4=x^4z^2-y^3-1\}$.

\noindent 4) Exponents found (part 3): $\{ n_1=1, n_2=0, n_3=0, n_4=2\}$. So, ${\cal P}_3=2{x}^{5}z \left( 2{x}^{4}{z}^{2}\!-\!2{y}^{3}\!-\!3 \right)$, ${\cal P}_2=-3x{y}^{2} \left( 2{x}^{4}{z}^{2}\!-\!2{y}^{3}\!-\!3 \right)$, ${\cal P}_1\!=\!10{x}^{8}{z}^{4}\!-\!12{x}^{4}{y}^{3}{z}^{2}\!-\!16{x}^{4}{z}^{2}$  $+2\,{y}^{6}+4\,{y}^{3}+2$.

\medskip

\noindent
The integrating factor and the LFI are given by
\begin{equation}
\label{s4ex2RandI}
R = \frac{{\rm e}^{\frac{-1}{{x}^{4}{z}^{2}-{y}^{3}-1}}}{\left( {x}^{4}{z}^{2}-{y}^{3}-1 \right) ^{2}{x}^{2}}, \,\,\,\,\,I ={\frac {{{\rm e}^{ \left( -{x}^{4}{z}^{2}+{y}^{3}+1 \right) ^{-1}}}y}{x}}+{\it Ei} \left( 1,- \left( -{x}^{4}{z}^{2}+{y}^{3}+1 \right) ^{-1} \right).
\end{equation}

\bigskip

\noindent {\bf 2ODE-3:}
\begin{eqnarray}
z' \!\!\!&=& \!\!\!\left(16\,{x}^{5}{y}^{7}{z}^{11}\!-\!16{x}^{4}{y}^{8}{z}^{10}\!+4{x}^{3}{y}^{9}{z}^{9}\!+\!{x}^{3}{y}^{8}{z}^{9}
+32\,{x}^{4}{y}^{3}{z}^{7}+8\,{x}^{4}{y}^{3}{z}^{6}-40\,{x}^{3}{y}^{4}{z}^{6} \right.
\nonumber \\ [2mm]
&& 
\left. \!\!\!\!\!\!\!\!-4{x}^{3}{y}^{4}{z}^{5}\!+\!16{x}^{2}{y}^{5}{z}^{5}\!+\!4{x}^{2}{y}^{4}{z}^{5}\!-\!2x{y}^{6}{z}^{4}
\!-\!16{x}^{2}{z}^{2}\!-\!4{x}^{2}z\!+\!16xyz\!+\!2xy\!+\!4xz\!-\!4{y}^{2}\right) 
\nonumber \\ [2mm]
&& 
\!\!\!\!\!\!\!/\!\left(-2{x}^{2}\left(  8\,{x}^{3}{y}^{8}{z}^{9}\!-\!8{x}^{2}{y}^{9}{z}^{8}\!+\!{x}^{2}{y}^{8}{z}^{8}
\!+\!2x{y}^{10}{z}^{7}\!+\!16{x}^{2}{y}^{4}{z}^{5}\!+\!4{x}^{2}{y}^{4}{z}^{4}\!-\!16x{y}^{5}{z}^{4} \right. \right.
\nonumber \\ [2mm]
&& 
\left. \left. \!\!\!\!\!\!\!\!-\!2x{y}^{5}{z}^{3}\!+\!4x{y}^{4}{z}^{4}+4\,{y}^{6}{z}^{3}+4 \right) \right).
\label{s4ex3}
\end{eqnarray}

\begin{table}[h]
{\begin{center} {\footnotesize
\begin{tabular}
{|c|c|c|c|}
\hline
Algorithm & Task & Memory (MB) & Time (sec)  \\
\hline
 {\it NLS} & $\{P_3,P_2,P_1\}$ & 1 & 0.063    \\
\hline
 {\it PLDIF} (part 1) & Coeff. reduction & 4 & 1.750   \\
\hline
 {\it PLDIF} (part 2) & Compute DPs & 0 & 0.202   \\
\hline
 {\it PLDIF} (part 3) & Compute $\Upsilon$ & 0 & 0.000   \\
\hline
 {\it NLSPLDIF} & Find a LFI $I$ & 5 & 2.015   \\
\hline
\end{tabular} }
\caption{Time and Memory consumption (one way) --- 2ODE (\ref{s4ex2})}
\end{center}}
\label{tabex3}
\end{table}

\begin{table}[h]
{\begin{center} {\footnotesize
\begin{tabular}
{|c|c|c|c|}
\hline
Algorithm & Task & Memory (MB) & Time (sec)  \\
\hline
 {\it NLS} & $\{P_3,P_2,P_1\}$ & 1 & 0.063    \\
\hline
 {\it PLDIF} (part 1) & Coeff. reduction & 17 & 10.500   \\
\hline
 {\it PLDIF} (part 2) & Compute DPs & 1 & 3.281   \\
\hline
 {\it PLDIF} (part 3) & Compute $\Upsilon$ & 0 & 0.000   \\
\hline
 {\it NLSPLDIF} & Find a LFI $I$ & 19 & 13.844   \\
\hline
\end{tabular} }
\caption{Time and Memory consumption (another way) --- 2ODE (\ref{s4ex2})}
\end{center}}
\label{tabex3-2}
\end{table}

\noindent
{\bf Procedure {\it NLS}:} 

\noindent 1) {\it NLS} computes $P_2 \!=\! \left( 16\,{x}^{4}{y}^{7}{z}^{10}-16\,{x}^{3}{y}^{8}{z}^{9}+4\,{x}^{2
}{y}^{9}{z}^{8}-{x}^{2}{y}^{8}{z}^{8}+32\,{x}^{3}{y}^{3}{z}^{6}\right.$

\noindent
$\left. +8\,{x}^{3}{y}^{3}{z}^{5}-32\,{x}^{2}{y}^{4}{z}^{5}-4\,{x}^{2}{y}^{4}{z}^{4}+8\,x{y}^{5}{z}^{4}-4\,x{y}^{4}{z}^{4}-4 \right)\! x$ and $P_1=2\, \left( 8\,{x}^{3}{y}^{8}{z}^{9}\right.$

\noindent
$\left. -8\,{x}^{2}{y}^{9}{z}^{8}+{x}^{2}{y}^{8}{z}^{8}\!+2\,x{y}^{10}{z}^{7}\!+16\,{x}^{2}{y}^{4}{z}^{5}\!+4\,{x}^{2}{
y}^{4}{z}^{4}\!-16\,x{y}^{5}{z}^{4}\!-2\,x{y}^{5}{z}^{3}\!+4\,x{y}^{4}{z}^{4}\right.$

\noindent
$\left.+4\,{y}^{6}{z}^{3}+4 \right) {x}^{2}$.

\medskip
\noindent
{\bf Procedure {\it PLDIF}:} 

\noindent 2) Reduction of coefficients (part 1 -- one way): 1575 $\rightarrow$ 4  and  3083 $\rightarrow$ 121

\noindent 2) Reduction of coefficients (part 1 -- another way): 4851 $\rightarrow$ 260

\noindent 3) DPs found (part 2): $\left\{ 2\,xz-y,x{y}^{4}{z}^{4}+2 \right\}$.

\noindent 4) Exponents found (part 3): $\{ n_1=1, n_2=2\}$. So, ${\cal P}_3=-2\,x \left( 5\,{x}^{2}{y}^{8}{z}^{8}-2\,x{y}^{9}{z}^{7}\right.$

\noindent
$\left. +2\,{x}^{2}{y}^{4}{z}^{4}-x{y}^{5}{z}^{3}+12\,x{y}^{4}{z}^{4}-4\,{y}^{5}{z}^{3}+4 \right)$, ${\cal P}_2=-8\,{x}^{3}{y}^{7}{z}^{9}+5\,{x}^{2}{y}^{8}{z}^{8}-4\,{x}^{3}{y}^{3}{z}^{5}$

\noindent
$+2\,{x}^{2}{y}^{4}{z}^{4}-16\,{x}^{2}{y}^{3}{z}^{5}+12\,x{y}^{4}{
z}^{4}+4$, ${\cal P}_1\!=-4\,{x}^{2}{y}^{8}{z}^{9}+x{y}^{9}{z}^{8}\!-12\,x{y}^{4}{z}^{5}\!+2\,{y}^{
5}{z}^{4}\!+2\,xz-y-8\,z$.

\medskip

\noindent
The integrating factor and the LFI are given by
\begin{equation}
\label{s4ex3RandI}
R = \frac{{\rm e}^{{\frac {x}{x{y}^{4}{z}^{4}+2}}}}{ \left( x{y}^{4}{z}^{4}+2
 \right) ^{2} \left( 2\,xz-y \right) ^{2}}, \,\,\,\,\,I ={{\rm e}^{{\frac {x}{x{y}^{4}{z}^{4}+2}}}} \left( -2\,xz+y \right) ^{-
1}+{\it Ei} \left( 1,-{\frac {x}{x{y}^{4}{z}^{4}+2}} \right).
\end{equation}

\medskip

\noindent {\bf 2ODE-4:}
\begin{equation}
\!\!\!z'\! =\!\frac {3{y}^{2}{z}^{5}\!\!+\!3{y}^{5}z\!+\!2y{z}^{5}\!\!-\!9{x}^{2}{y}^{2}z\!+\!2x{z}^{4}
\!-\!{y}^{4}z\!-\!9{y}^{2}{z}^{3}\!-\!2{x}^{2}yz\!+\!2x{y}^{3}\!+\!2{x}^{3}\!+\!2x{y}^{2}\!+\!6x{z}^{2}\!\!}{2z \left( -2\,{z}^{6}-2\,{y}^{3}{z}^
{2}+6\,{x}^{2}{z}^{2}+2\,{y}^{2}{z}^{2}+3\,{z}^{4}-3\,{y}^{3}+3\,{x}^{2} \right)}\!.
\label{s4ex4}
\end{equation}

\begin{table}[h]
{\begin{center} {\footnotesize
\begin{tabular}
{|c|c|c|c|}
\hline
Algorithm & Task & Memory (MB) & Time (sec)  \\
\hline
 {\it NLS} & $\{P_3,P_2,P_1\}$ & 1 & 0.031    \\
\hline
 {\it PLDIF} (part 1) & Coeff. reduction & 0 & 0.032   \\
\hline
 {\it PLDIF} (part 2) & Compute DPs & 0 & 0.016   \\
\hline
 {\it PLDIF} (part 3) & Compute $\Upsilon$ & 0 & 0.000   \\
\hline
 {\it NLSPLDIF} & Find a LFI $I$ & 1 & 0.079   \\
\hline
\end{tabular} }
\caption{Time and Memory consumption --- 2ODE (\ref{s4ex4})}
\end{center}}
\label{tabex4}
\end{table}

\noindent
{\bf Procedure {\it NLS}:} 

\noindent 1)$\!\!$ {\it NLS} computes$\!$ $P_2 \!=\! xz \left( {x}^{2}yz-3\,xy{z}^{2}-2\,x+6\,z-1 \right)  \left( x-3\,z
 \right)$ and $P_1={x}^{3}{y}^{2}{z}^{2}-6\,{x}^{2}{y}^{2}{z}^{3}+9\,x{y}^{2}{z}^{4}-{x}^
{2}{y}^{2}{z}^{2}-2\,{x}^{2}yz+12\,xy{z}^{2}-18\,y{z}^{3}+3\,xyz+3\,y{z}^{2}-4$.

\medskip
\noindent
{\bf Procedure {\it PLDIF}:} 

\noindent 2) Reduction of coefficients (part 1): 203 $\rightarrow$ 11

\noindent 3) DPs found (part 2): $\{ p_1=x, p_2=y, p_3=z, p_4=-{z}^{4}-{y}^{3}+{x}^{2}\}$.

\noindent 4) Exponents of $R$ found directly (elementary first integral): $\{ n_1=0, n_2=0, n_3=0, n_4=2\}$.

\medskip

\noindent
The integrating factor and the LFI are given by
\begin{equation}
\label{s4ex4RandI}
R = \frac{1}{\left(-{z}^{4}-{y}^{3}+{x}^{2} \right) ^{2}}, \,\,\,\,\,I =\frac{{{\rm e}^{{\frac {2\,{x}^{2}+{y}^{2}+3\,{z}^{2}}{-{z}^{4}-{y}^{3}+{x}^
{2}}}}}}{ \left( -{z}^{4}-{y}^{3}+{x}^{2} \right)}.
\end{equation}

\medskip
\subsubsection{Second set:}

\noindent {\bf 2ODE-5:}
\begin{eqnarray}
z' \!\!\!\!\!&=& \!\!\!\!\!- \left({x}^{4}y{z}^{3}\!-\!6{x}^{3}y{z}^{4}\!+\!9{x}^{2}y{z}^{5}\!+\!{x}^{3}{y}^{2}{z}^{2}
\!-\!6{x}^{2}{y}^{2}{z}^{3}\!+\!9x{y}^{2}{z}^{4}\!-\!{x}^{2}{y}^{2}{z}^{2}\!-\!2{x}^{3}{z}^{2}\!+\!12{x}^{2}{z}^{3} \right.
\nonumber \\ [2mm]
&& 
\left. \!\!\!\!\!\!\!\!\!\!-18x{z}^{4}\!-\!2{x}^{2}yz\!-\!{x}^{2}{z}^{2}\!+\!12xy{z}^{2}\!+\!3x{z}^{3}\!-\!18y{z}^{3}
\!+\!3xyz\!+\!3y{z}^{2}\!-\!4\right)\! /\!\left({x}^{4}{y}^{2}z\!-\!6{x}^{3}{y}^{2}{z}^{2} \right.
\nonumber \\ [2mm]
&& 
\left. \!\!\!\!\!\!\!\!\!+9{x}^{2}{y}^{2}{z}^{3}\!+\!3{x}^{2}{y}^{2}{z}^{2}\!-\!2{x}^{3}y\!+\!12{x}^{2}yz\!-\!18xy{z}^{2}
\!-\!{x}^{2}y\!-\!9xyz+\!12 \right).
\label{s4ex5}
\end{eqnarray}

\begin{table}[h]
{\begin{center} {\footnotesize
\begin{tabular}
{|c|c|c|c|}
\hline
Algorithm & Task & Memory (MB) & Time (sec)  \\
\hline
 {\it NLS} & $\{P_3,P_2,P_1\}$ & 1 & 0.078    \\
\hline
 {\it PLDIF} (part 1) & Coeff. reduction & 1 & 0.046   \\
\hline
 {\it PLDIF} (part 2) & Compute DPs & 0 & 0.015   \\
\hline
 {\it PLDIF} (part 3) & Compute $\Upsilon$ & 4 & 0.563   \\
\hline
 {\it NLSPLDIF} & Find a LFI $I$ & 6 & 0.703   \\
\hline
\end{tabular} }
\caption{Time and Memory consumption --- 2ODE (\ref{s4ex5})}
\end{center}}
\label{tabex5}
\end{table}

\noindent
{\bf Procedure {\it NLS}:} 

\noindent 1)$\!\!$ {\it NLS} computes$\!$ $P_2\!=\! xz \left( {x}^{2}yz-3\,xy{z}^{2}-2\,x+6\,z-1 \right)  \left( x-3\,z
 \right)$ and $P_1={x}^{3}{y}^{2}{z}^{2}-6\,{x}^{2}{y}^{2}{z}^{3}+9\,x{y}^{2}{z}^{4}-{x}^
{2}{y}^{2}{z}^{2}-2\,{x}^{2}yz+12\,xy{z}^{2}-18\,y{z}^{3}+3\,xyz+3\,y{z}^{2}-4$.

\medskip
\noindent
{\bf Procedure {\it PLDIF}:} 

\noindent 2) Reduction of coefficients (part 1): 168 $\rightarrow$ 5;  243 $\rightarrow$ 21

\noindent 3) DPs found (part 2): $\{ p_1=x, p_2=y, p_3=x-3z, p_4=xyz-2 \}$.

\noindent 4) Exponents of $R$ found directly (elementary first integral): $\{ n_1=0, n_2=0, n_3=1, n_4=2\}$. 

\medskip

\noindent
The integrating factor and the LFI are given by
\begin{equation}
\label{s4ex5RandI}
R = \frac{1}{\left(-{z}^{4}-{y}^{3}+{x}^{2} \right) ^{2}}\,, \,\,\,\,\,\,I =\frac{{{\rm e}^{{\frac {2\,{x}^{2}+{y}^{2}+3\,{z}^{2}}{-{z}^{4}-{y}^{3}+{x}^
{2}}}}}}{ \left( -{z}^{4}-{y}^{3}+{x}^{2} \right)}.
\end{equation}

\medskip

\noindent {\bf 2ODE-6:}
\begin{equation}
\!\!\!z'\! =\!{\frac {-{x}^{3}{z}^{4}\!+\!4{z}^{7}\!+\!{x}^{4}yz\!-\!{x}^{2}y{z}^{3}\!-\!8xy{z}^
{4}\!+\!{x}^{3}{y}^{2}\!+\!4{x}^{2}{y}^{2}z\!-\!{x}^{3}z\!+\!4{x}^{2}{z}^{2}\!-\!{x}^{
2}y\!+\!4xyz}{x \left( -3\,x{z}^{5}+4\,{z}^{6}+3\,{x}^{2}y{z}^{2}-8\,xy{
z}^{3}+4\,{x}^{2}{y}^{2}-3\,x{z}^{2}+12\,{z}^{3} \right) }}\!.
\label{s4ex6}
\end{equation}

\begin{table}[h]
{\begin{center} {\footnotesize
\begin{tabular}
{|c|c|c|c|}
\hline
Algorithm & Task & Memory (MB) & Time (sec)  \\
\hline
 {\it NLS} & $\{P_3,P_2,P_1\}$ & 1 & 0.000    \\
\hline
 {\it PLDIF} (part 1) & Coeff. reduction & 1 & 0.016   \\
\hline
 {\it PLDIF} (part 2) & Compute DPs & 0 & 0.047   \\
\hline
 {\it PLDIF} (part 3) & Compute $\Upsilon$ & 0 & 0.109   \\
\hline
 {\it NLSPLDIF} & Find a LFI $I$ & 2 & 0.172   \\
\hline
\end{tabular} }
\caption{Time and Memory consumption --- 2ODE (\ref{s4ex6})}
\end{center}}
\label{tabex6}
\end{table}

\noindent
{\bf Procedure {\it NLS}:} 

\noindent 1) {\it NLS} computes $P_2 \!=\! - \left( -x{z}^{3}+{x}^{2}y-x+4\,z \right) {x}^{2}$ and $P_1=-4\,{z}^{7}+{x}^{2}y{z}^{3}+8\,xy{z}^{4}-{x}^{3}{y}^{2}-4\,{x}^{2}{y}^{2}z+{x}^{2}y-4\,xyz$.

\medskip
\noindent
{\bf Procedure {\it PLDIF}:} 

\noindent 2) Reduction of coefficients: 288 $\rightarrow$ 24

\noindent 3) DPs found (part 2): $\left\{ x, -{z}^{3}+xy \right\}$.

\noindent 4) Exponents found (part 3): $\{ n_1=1, n_2=2\}$. So, ${\cal P}_3=-3\,x \left( -2\,{z}^{3}+2\,xy+1 \right) {z}^{2}
$, ${\cal P}_2={x}^{2} \left( -2\,{z}^{3}+2\,xy+1 \right)$, ${\cal P}_1\!=2\,{z}^{6}-6\,xy{z}^{3}+4\,{x}^{2}{y}^{2}+xy$.

\medskip

\noindent
The integrating factor and the LFI are given by
\begin{equation}
\label{s4ex6RandI}
R = \frac{{\rm e}^{{\frac {x}{x{y}^{4}{z}^{4}+2}}}}{ \left( x{y}^{4}{z}^{4}+2
 \right) ^{2} \left( 2\,xz-y \right) ^{2}}, \,\,\,\,\,I =\frac{{\rm e}^{{\frac {x}{x{y}^{4}{z}^{4}+2}}}}{\left( -2\,xz+y \right)}+{\it Ei} \left( 1,-{\frac {x}{x{y}^{4}{z}^{4}+2}} \right).
\end{equation}

\medskip

\noindent {\bf 2ODE-7:}
\begin{equation}
\!\!\!\!z'\! =\!{\frac {2{x}^{4}z\!-\!2{x}^{3}{z}^{2}\!-\!2{x}^{2}y{z}^{2}\!+\!2xy{z}^{3}
\!-\!2{x}^{2}yz\!-\!2x{z}^{3}\!+\!2{y}^{2}{z}^{2}\!+\!y{z}^{3}+{z}^{4}\!-\!2{x}^{3
}\!+\!4{x}^{2}z\!-\!2x{z}^{2}}{-\left(2\,{x}^{3}y+2\,{x}^{3}z-{x}^{2}{z}^{2}-2\,x
{y}^{2}z-2\,xy{z}^{2}+y{z}^{3}+{x}^{2}y-2\,xyz+y{z}^{2}\right)}}\!.
\label{s4ex7}
\end{equation}

\begin{table}[h]
{\begin{center} {\footnotesize
\begin{tabular}
{|c|c|c|c|}
\hline
Algorithm & Task & Memory (MB) & Time (sec)  \\
\hline
 {\it NLS} & $\{P_3,P_2,P_1\}$ & 1 & 0.047    \\
\hline
 {\it PLDIF} (part 1) & Coeff. reduction & 0 & 0.015   \\
\hline
 {\it PLDIF} (part 2) & Compute DPs & 0 & 0.078   \\
\hline
 {\it PLDIF} (part 3) & Compute $\Upsilon$ & 0 & 0.000   \\
\hline
 {\it NLSPLDIF} & Find a LFI $I$ & 1 & 0.140   \\
\hline
\end{tabular} }
\caption{Time and Memory consumption --- 2ODE (\ref{s4ex7})}
\end{center}}
\label{tabex7}
\end{table}

\noindent
{\bf Procedure {\it NLS}:} 

\noindent 1) {\it NLS} computes $P _2\!=\! \left( 2\,{x}^{3}-2\,xyz+xz-{z}^{2} \right)  \left( x-z \right)$ and $P_1=-2\,{x}^{2}yz-{x}^{2}{z}^{2}+2\,{y}^{2}{z}^{2}+y{z}^{3}-2\,{x}^{3}+4\,{x}^{2}z-2\,x{z}^{2}$.

\medskip
\noindent
{\bf Procedure {\it PLDIF}:} 

\noindent 2) Reduction of coefficients: 126 $\rightarrow$ 26

\noindent 3) DPs found (part 2): $\left\{ x-z, x^2-yz \right\}$.

\noindent 4) Exponents found (part 3): $\{ n_1=1, n_2=2\}$. So, ${\cal P}_3=-2\,{x}^{3}y-2\,{x}^{3}z+{x}^{2}{z}^{2}+2\,x{y}^{2}z+2\,xy{z}^{2}-y{z}
^{3}-2\,{x}^{3}-2\,{x}^{2}y+2\,{x}^{2}z+6\,xyz-4\,y{z}^{2}$, ${\cal P}_1\!=2\,{x}^{2}yz+{x}^{2}{z}^{2}-2\,{y}^{2}{z}^{2}-y{z}^{3}+6\,{x}^{3}-10\,{x}^{2}z-2\,xyz+4\,x{z}^{2}+2\,y{z}^{2}$, ${\cal P}_2=-2\, \left( x-z \right)  \left( {x}^{3}-xyz+xz-{z}^{2} \right)$.

\medskip

\noindent
The integrating factor and the LFI are given by
\begin{equation}
\label{s4ex7RandI}
R = \frac{{\rm e}^{{\frac {2\,xy+{z}^{2}}{x-z}}}}{\left( x-z \right) ^{2}
 \left( {x}^{2}-zy \right) ^{2}}, \,\,\,\,\,I =\frac {2\,xy+{z}^{2}}{x-z}-\ln\left( {x}^{2}-zy \right).
\end{equation}

\medskip

\subsection{Some final observations and possible developments}
\label{fincon}

Although the 2ODEs presented in the previous subsection establish a first (summarized) analysis of the efficiency of the developed algorithms, several points still need to be raised/studied. In this section, we will highlight some of these points and briefly discuss possible ways forward in this line:

\begin{enumerate}
\item In all examples we used only one of the equations ($E_1$,$E_2$,$E_3$). In 2ODE (\ref{s4ex3}) we present the difference in the result when using one or other of these equations.

\item The difference in CPU time and memory consumption in 2ODE (\ref{s4ex3}) is not common. In general, CPU time and memory consumption are very similar for the three equations ($E_1$,$E_2$,$E_3$).

\item For the first set (2ODEs 1 to 4) the S-function method is unable to determine the symmetry in a short time ($\leq$ 30 seconds). Even if the symmetry was found, the associated 1ODE can not be solved by the  {\tt dsolve} command of the CAS Maple. For the second set the time/memory costs are in table 9. 

Obs.: In table 9, the symbol (*) means that {\tt dsolve} was not able to solve the associated 1ODEs, that is, only the symmetry calculation was performed.

\begin{table}[h]
{\begin{center} {\footnotesize
\begin{tabular}
{|c|c|c|}
\hline
2ODE & Memory (MB) & Time (sec)  \\
\hline
5 & 19 & 5.047 (*)   \\
\hline
6 & 4 & 0.956 (*)   \\
\hline
7 & 3 & 0.235   \\
\hline
\end{tabular} }
\caption{S-function Method --- 2ODEs}
\end{center}}
\label{tabex8}
\end{table}

\item The determination of the multiplicities of the Darboux polynomials that are factors of $\Upsilon$ (part 3 of the {\it PLDIF} procedure) is only necessary when the 2ODE presents a non-elementary LFI. However, even in these cases, the multiplicities are very often provided by the probabilistic algorithm itself. For instance, in example \ref{eximp1} the probabilistic algorithm (using $\mathfrak{X}_3$) returns the following DPs: $\left\{ {x}, \left( xy-z \right) ^{2},xy-z,{x}^{2}z-y \right\}$. In that example $\Upsilon= \left( xy-z \right) ^{2} \left( {x}^{2}z-y \right) x$ and so, we didn't have to iterate the $\Upsilon_c$ candidate with the known Darboux polynomials.

\bigskip

\item There are some questions linked to the structure of the integrating factors and first integrals that, at this stage of the study, have not yet been answered. Some of the main questions are:
\begin{enumerate}
\item So far we have not been able to establish a bound for the degree of the polynomials ${\cal P}_1,\,{\cal P}_2,\,{\cal P}_3,\,\Upsilon$. 

\medskip

{\bf Question 1:} In the `non-degenerate' case ($\Upsilon$ is a polynomial), can we establish an upper bound on the degree of the polynomials ${\cal P}_1,\,{\cal P}_2,\,{\cal P}_3,\,\Upsilon$? 
\begin{obs}
In affirmative case this would impose a bound on the degree of the Darboux polynomials that are factors of $\Upsilon$.
\end{obs}

\item The space of solutions ${\cal V}$ of the linear system of indeterminates $S_{E} \equiv \bigcup_i S_{E_i}$ is a linear space of a certain dimension $d_s$.

\medskip

{\bf Question 2:} How is $d_s$ related to the vector field $\mathfrak{X}$? 

\medskip
Sub-questions: How are the dimensions ${d_s}_i$ of the linear spaces ${\cal V}_i$ (${\cal V}_i\,\equiv$ the space of solutions of the system $S_{E_i}$) related to the vector fields $\mathfrak{X}_i$? Is $d_s$ related to the number of irreducible Darboux polynomials present in the integrating factor and/or to the number of exponential factors?

\item The polynomials that appear multiplied by the remaining coefficients when we substitute the solution $S_{sys}$ in $\Upsilon_c$ are the basis vectors of a possible representation of the linear space ${\cal V}$. In example \ref{eximp1} we saw that one of these `basis vectors' was the very solution we were looking for ($\Upsilon$).

\medskip
{\bf Question 3:} Is there a `canonical' basis \mbox{\boldmath $e$}$_{\cal V}$ for the space ${\cal V}$, in which the sought solution for $\Upsilon$ is one of the basis vectors in the representation of ${\Upsilon_c}$'s partial solution? 

\medskip
Sub-questions: In \mbox{\boldmath $e$}$_{\cal V}$, what do the other basis vectors can represent? Is it possible to compute \mbox{\boldmath $e$}$_{\cal V}$ in a linear way without using the probabilistic algorithm {\it PLDIF}?
\end{enumerate}

\item The idea of vector fields that `share' the Darboux polynomials seems to be very fruitful and there is still much to be studied and improved. One of the main virtues of this type of idea is that it seems to be generalizable to ODEs of order higher than two and, possibly, to partial differential equations. 

\end{enumerate}

%%%%%%%%%%%%%%%%%%%%%%%%%%%%%%

\end{document}